%% file: duality25oct.tex
\documentclass[a4paper,11pt]{article}
\usepackage{graphicx,rotating,hyperref,slashed,amsmath,xcolor,amssymb,amsfonts,expdlist,colortbl,cite,expdlist}
\pdfoutput=1
\makeatletter
\hypersetup{colorlinks,bookmarksopen,bookmarksnumbered,
linkcolor=blus,pdfstartview=FitH,urlcolor=rossos,citecolor=verde}
\allowdisplaybreaks

\def\lsim{\mathrel{\rlap{\lower3pt\hbox{\hskip0pt$\sim$}}
   \raise1pt\hbox{$<$}}}         
\def\gsim{\mathrel{\rlap{\lower4pt\hbox{\hskip1pt$\sim$}}
   \raise1pt\hbox{$>$}}}         

 \newcommand{\sfootnote}[1]{}
\definecolor{bluc}{cmyk}{1,1,0,0.1}
\definecolor{rossoCP3}{cmyk}{0,.88,.77,.40}
\definecolor{rosso}{cmyk}{0,1,1,0.4}
\definecolor{rossos}{cmyk}{0,1,1,0.55}
\definecolor{rossoc}{cmyk}{0,1,1,0.2}
\definecolor{verdes}{cmyk}{0.92,0,0.59,0.4}

\hypersetup{colorlinks, bookmarksopen, bookmarksnumbered,
citecolor=verdes, linkcolor=bluc, pdfstartview=FitH, urlcolor=rossos}

\newcommand{\mio}[1]{}

\definecolor{Gray}{gray}{0.95}

\usepackage{multicol}
\usepackage{color}
\definecolor{rosso}{cmyk}{0,1,1,0.4}
\definecolor{rossos}{cmyk}{0,1,1,0.55}
\definecolor{rossoc}{cmyk}{0,1,1,0.2}
\definecolor{blu}{cmyk}{1,1,0,0.3}
\definecolor{blus}{cmyk}{1,1,0,0.6}
\definecolor{bluc}{cmyk}{1,1,0,0.1}
\definecolor{verde}{cmyk}{0.92,0,0.59,0.25}
\definecolor{verdec}{cmyk}{0.92,0,0.59,0.15}
\definecolor{verdes}{cmyk}{0.92,0,0.59,0.4}

\def\circa#1{\,\raise.3ex\hbox{$#1$\kern-.75em\lower1ex\hbox{$\sim$}}\,}

\newcommand{\beq}{\begin{equation}}
\newcommand{\eeq}{\end{equation}}

\newcommand{\bea}{\begin{eqnarray}}
\newcommand{\eea}{\end{eqnarray}}
\newcommand{\be}{\begin{equation}}
\newcommand{\ee}{\end{equation}}
\newfam\rsfsfam
\def\mathscr#1{{\fam\rsfsfam\relax#1}}

\def\circa#1{\,\raise.3ex\hbox{$#1$\kern-.75em\lower1ex\hbox{$\sim$}}\,}
\makeatletter

\def\hhref#1{\href{http://arxiv.org/abs/#1}{arXiv:#1}} 

\newcommand{\doi}[1]{\href{http://dx.doi.org/#1}{[doi]}}

\def\hhref#1{\href{http://arxiv.org/abs/#1}{arXiv:#1}} 
 
\def\art{\@ifnextchar[{\eart}{\oart}}
\def\eart[#1]#2#3#4#5#6{{\rm #2}, {\em #3 \bf #4} {\rm (#6) #5} ({\em #1})}

\def\article{\@ifnextchar[{\earticle}{\oarticle}}
\def\oarticle#1#2#3#4#5#6{{\rm #1}, {\em ``#6''}, {\rm #2 #3 (#5) #4}}
\def\earticle[#1]#2#3#4#5#6#7{{\rm #2}, {\em ``#7''}, {\rm #3 #4 (#6) #5}  [\hhref{#1}]}
\def\hepart[#1]#2{{\rm #2, \em#1}}
\def\heparticle[#1]#2#3{#2, {\em ``#3''} [\hhref{#1}]}

\newcounter{alphaequation}[equation]
\def\thealphaequation{\theequation\hbox to
0.6em{\hfil\alph{alphaequation}\hfil}}
\def\eqnsystem#1{
\def\@eqnnum{{\rm (\thealphaequation)}}
\def\@@eqncr{\let\@tempa\relax \ifcase\@eqcnt \def\@tempa{& & &} \or
  \def\@tempa{& &}\or \def\@tempa{&}\fi\@tempa
  \if@eqnsw\@eqnnum\refstepcounter{alphaequation}\fi
\global\@eqnswtrue\global\@eqcnt=0\cr}
\refstepcounter{equation} \let\@currentlabel\theequation \def\@tempb{#1}
\ifx\@tempb\empty\else\label{#1}\fi
\refstepcounter{alphaequation}
\let\@currentlabel\thealphaequation
\global\@eqnswtrue\global\@eqcnt=0 \tabskip\@centering\let\\=\@eqncr
$$\halign to \displaywidth\bgroup \@eqnsel\hskip\@centering
$\displaystyle\tabskip\z@{##}$&\global\@eqcnt\@ne
\hskip2\arraycolsep\hfil${##}$\hfil& \global\@eqcnt\tw@\hskip2\arraycolsep
$\displaystyle\tabskip\z@{##}$\hfil
\tabskip\@centering&\llap{##}\tabskip\z@\cr}
\def\endeqnsystem{\@@eqncr\egroup$$\global\@ignoretrue} \makeatother

\definecolor{fiorentina}{rgb}{.5,0,.5}

\begin{document}

\vspace{1truecm}
 
\begin{center}
\boldmath

{\textbf{\Large A geometrical approach to  degenerate scalar-tensor theories}}
\unboldmath

\bigskip\bigskip

\vspace{0.1truecm}

{  Javier Chagoya, Gianmassimo Tasinato}
 \\[5mm]
{\it Department of Physics, Swansea University, Swansea, SA2 8PP, U.K.}

\vspace{0.6cm}

\thispagestyle{empty}
\begin{quote}
 Degenerate scalar-tensor theories are recently proposed covariant theories of gravity coupled with a scalar field. Despite being characterised by higher order equations of motion, they do not propagate more than three degrees of freedom, thanks to the existence of constraints.
We discuss a geometrical approach to degenerate scalar-tensor systems, and analyse   its consequences. We show that some of these theories emerge as a certain limit of DBI Galileons. In absence of dynamical gravity, these systems correspond to scalar theories enjoying a symmetry which is different from Galileon invariance. The scalar theories have however problems concerning the propagation of fluctuations around a time dependent background. These issues can be tamed by breaking the symmetry by hand, or by minimally coupling the scalar with dynamical  gravity in a way that leads to  degenerate scalar-tensor systems. We show that distinct theories can be connected by a relation which generalizes Galileon duality, in certain cases also when gravity is dynamical. We discuss some implications of our results in concrete examples. Our findings can be helpful for assessing stability properties and understanding the non-perturbative structure of systems based on degenerate scalar-tensor systems.
 \end{quote}
\thispagestyle{empty}
\end{center}

\setcounter{page}{1}
\setcounter{footnote}{0}



\section{Introduction}

Galileons are a  class of scalar theories with derivative self interactions, characterised by equations of motion (EOMs) of second order, 
and satisfying a symmetry transformation $\phi\to \phi+a +b_\mu x^\mu$ \cite{Nicolis:2008in}. They have several features  motivating  their study: 
their structure is stable under loop corrections  thanks to non-renormalization theorems \cite{Luty:2003vm,Nicolis:2004qq} and  is 
closed under a duality \cite{deRham:2013hsa}.  
 Their S-matrix has special, distinctive properties \cite{Kampf:2014rka,Cheung:2014dqa,Cachazo:2014xea}. They exhibit superluminal behaviour around certain
 sources, possibly providing  consistent theoretical set-ups  to study such phenomenon.  
  When coupled with gravity,  Galileon symmetry is normally broken; on the other hand, it is possible to covariantize
Galileons in such a way that they maintain second order EOMs \cite{Deffayet:2009mn,Deffayet:2009wt}. Covariant Galileons  are especially interesting  for their cosmological
applications: they can lead to self accelerating cosmologies, and at the same time `hide' the presence of a scalar fifth
force  against local measurements
of gravitational interactions,  through an efficient Vainshtein screening mechanism.  
 See \cite{Clifton:2011jh,Joyce:2014kja} for reviews. Galileons and their covariantized versions 
have a geometrical interpretation, arising as certain limits of the so  called DBI Galileons, 
 which describe
  the features of probe branes embedded in an extra dimensional set-up \cite{deRham:2010eu,Goon:2011uw,Goon:2011qf}.  

The fact that Galilean symmetry is normally broken when coupling Galileons with gravity does not necessarily mean that the structure of the resulting actions
is not protected: in certain situations, gravitational interactions can break Galilean symmetries in a soft way, yet preserving some of the
desired features of Galileons. See e.g. \cite{Burrage:2010cu,Pirtskhalava:2015nla,Gratia:2016tgq}
 for examples on this respect. When coupled with gravity demanding second order EOMs, Galileons
admit as maximal extension the theories of Horndeski \cite{Horndeski:1974wa}, which  are the most general scalar-tensor theories characterised by 
 second order EOMs. 

Interestingly, it has been recently realized that
Horndeski   theories  are {\it not} the most general covariant scalar-tensor theories 
with  
at most three degrees of freedom (that is,  theories that  do not propagate additional ghostly modes). 
 Other possibilities  arise, when considering  degenerate scalar-tensor set-ups.
  The existence of  primary constraints   prevents the propagation of additional  degrees of freedom,
even for theories 
 characterised by 
equations of  motion of 
  order higher than two. 
   Examples  are the theories of beyond Horndeski (bH)  
\cite{Gleyzes:2014dya,Gleyzes:2014qga,Gao:2014soa}. Such systems have been recently further generalised to the so called Extended Scalar Tensor/(Degenerate Higher Order Scalar Tensor) theories, EST/(DHOST)  in \cite{Langlois:2015cwa,Crisostomi:2016czh,Achour:2016rkg}, using methods and
tools  developed in seminal papers by Langlois and Noui \cite{Langlois:2015cwa,Langlois:2015skt}.
Some of bH (or more generally EST) theories are known to be related with standard Horndeski Lagrangians
through conformal or disformal transformations; others, instead, seem to  lie at isolated points in the space of scalar-tensor theories  \cite{Crisostomi:2016tcp,Achour:2016rkg,deRham:2016wji}.  The study  of these  theories is still in its infancy, but by now we know that they can have consequences for cosmology and  screening mechanisms:  they lead to a breaking of the Vainshtein mechanism inside matter, modifying the internal structure of non relativistic stars \cite{Kobayashi:2014ida,Koyama:2015oma,Saito:2015fza,Sakstein:2015zoa,Jain:2015edg}.

The aim of this work is to address the following questions:
\begin{itemize}
\item[-] Is there some  relation between bH or EST actions and other  known scalar-tensor theories with a well understood  extra dimensional origin? In particular, are
  there any connections with extra dimensional models as Dirac-Born-Infeld  (DBI) and Dvali-Gabadadze-Porrati (DGP) set-ups? Addressing this question would allow one to 
     set these theories  in a broader context, and to apply to them  results and geometrical techniques  developed when studying other
     systems. 
\item[-] Is the structure  of  bH or EST theories protected by some symmetry, at least in certain cases? And is this
structure  closed
under a duality, as it happens for Galileon Lagrangians? 
This information can be important to examine the  stability of these theories under loop corrections, and to understand  
whether their distinctive non linear properties can be connected through  dualities to  features of simpler systems.
\end{itemize}
We do some preliminary steps towards answering    the previous points. It has been noticed already
 in \cite{Gleyzes:2014qga} that a particular choice of the free functions characterising  bH  Lagrangians give theories which,
 in absence of dynamical gravity, reduce to quartic and quintic Galileons. This implies that a naive covariantization of standard quartic and quintic Galileons
  (when expressed in the specific  form  $L_4^{\text{gal},1}$ and $L_5^{\text{gal},1}$ , using the nomenclature of \cite{Deffayet:2013lga}) is ghost-free.
   In absence of gravity,
  these bH theories acquire
  standard Galileon invariance. 
  
  Motivated by a construction in terms of a probe brane in an extra dimensional
  set-up,  we show here that {\it other} examples of theories belonging to the bH class provide set-ups
   which, with no dynamical
   gravity,  
   respect a symmetry
  different from Galileon symmetry. 
    Namely 
  \be\label{sym-int}
  \delta \pi\,=\,\pi \,\omega^\mu\,\partial_\mu \pi\,,
  \ee 
  with $\omega^\mu$ a constant vector. 
  Moreover, special cases of  bH Lagrangians at different orders  are  connected by a duality transformation, which generalizes the standard Galileon duality. 
  
 \smallskip
 We organise our work in successive steps, to build up  
 the tools necessary to discuss our results:
\begin{itemize}
\item We start in Section \ref{sec-intga} with a  review of a determinantal approach to standard Galileon Lagrangians, pointing out that it
is particularly convenient to make manifest how Galileon dualities connect different Galileon actions.
\item We then discuss in Section \ref{sec-dbigals} a novel perspective to DBI Galileons based on a determinantal approach to these systems. This
is different with respect  to the usual approach which obtains DBI Galileons starting from specific curvature invariants that 
form  the action for a brane 
   probing
extra dimensional space. Our point of view is convenient for discussing dualities among DBI Galileon Lagrangians, and  for making contact
with degenerate scalar-tensor theories when coupling with gravity.
\item Section \ref{sec-dbiextr} studies a novel, particular limit of DBI Galileons which we dub extreme relativistic.
 In order for the limit to be well defined, we have to consider extra dimensional space times with different signatures,
depending on whether the vector normal to the probe brane is time like or space like. 
 The extreme relativistic limit is 
   opposite to the non relativistic
limit of DBI Galileon actions, which provides standard Galileons.  
The resulting scalar theories  have peculiar features which we point out, and are characterised by
the field dependent symmetry of eq. \eqref{sym-int}. Moreover, scalar theories of different orders are again connected by dualities. The scalar theories have problems
since fluctuations around interesting backgrounds do not have proper kinetic terms: this issue
 can be tamed by weakly breaking the symmetry.  
\item Section \ref{sec-est} shows that a minimal covariantization of the DBI Galileons in the extreme relativistic limit gives consistent scalar-tensor theories of gravity, despite being 
characterized by equations of motion of order higher than two. Indeed, the resulting system corresponds to a particular case of beyond Horndeski Lagrangians. We also  discuss how to  further generalise our results to include extended scalar-tensor theories.  
This construction provides a geometrical perspective to degenerate scalar-tensor theories. 
The strong coupling problems we met  in Section  \ref{sec-dbiextr}  are automatically solved
 when the scalar theories are coupled with gravity. On the other hand,
   the scalar symmetry  gets usually broken, although  for
  certain cases some of its properties can be preserved. 
  Moreover,  we show that 
  in certain cases 
  different classes of 
  such 
   theories can be  connected by dualities, also when dynamical gravity is turned on.
 Our results  can be helpful
for assessing stability properties or understanding the non-perturbative structure of degenerate scalar-tensor systems.
\end{itemize}
We conclude in Section \ref{sec-disc} discussing  possible follow ups for our work, while two appendixes contain further technical details.

\section{Symmetries and dualities for  Galileon Lagrangians}\label{sec-intga}

To set the stage, in this Section
 we succinctly   review   basic properties of Galileon scalar theories in four dimensional flat space, in absence of gravity, 
 adopting a method   that will be useful for what comes next. The  use of 
  an approach based on the Levi Civita symbol, discussed for example in
 \cite{Curtright:2012gx}, is particularly suitable for studying dualities among the actions we consider, as well as for investigating their symmetries.

 \subsection{The Galileon system}

 Galileon theories are
 described by scalar actions which lead to equations of motion of second order, and
 satisfy a Galileon symmetry.  There are several ways to express Galileon actions (see for
 example the reviews \cite{Curtright:2012gx,Deffayet:2013lga}). Here, we adopt a determinantal form whose
 basic building block is the  scalar action
\be\label{actg1}
{\cal S} \,=\, {\cal N}\,\int \,d^4 x\,\left( \partial \pi \right)^2\,\det{\left[ {\bf 1}+ c \,{\bf \Pi}\right]}\,,
\ee
with ${\cal N}$ an overall normalization,   $c$ a dimensionful constant,  ${\bf 1}$ the unit matrix in four dimensions, and ${ \bf \Pi}$ the $4\times 4$ symmetric
 matrix whose components are 
$${\bf \Pi}\,=\,\Pi_\mu^\nu\,=\,\partial_\mu \partial^\nu\,\pi \,.$$ 

A determinantal form for action \eqref{actg1} is convenient once we recall the definition of 
  determinant of a matrix $\bf M$  in terms of the antisymmetric Levi Civita
symbol: 
\be
\det{\bf M}\,=\,\frac{1}{4!}\,M_{\mu_1}^{\nu_1}\,M_{\mu_2}^{\nu_2}\,M_{\mu_3}^{\nu_3}\,M_{\mu_4}^{\nu_4}\,\epsilon_{\nu_1\nu_2\nu_3\nu_4 }\,\epsilon^{\mu_1\mu_2\mu_3\mu_4 }\,.
\ee
This implies that
\bea
\det\left[{\bf 1}+{\bf M}\right]&=&
1+ \,\left[M \right]+\frac{1}{2}\left( 
\left[M\right]^2-\left[M^2\right]\right)
+\frac{1}{6}\left(
\left[M\right]^3-3\left[M \right]\left[M^2 \right]+2 \left[M^3 \right]
 \right)
 \nonumber\\
 &&
 +\frac{1}{24}\left(
\left[M\right]^4-6\left[M \right]^2\left[M^2 \right]
+3 \left[M^2 \right]^2
+8 \left[M \right]\left[M^3 \right]-6\left[M^4 \right]
 \right)
 \label{det-for} \eea
 where  $[M]\,= \,{\text{tr}}\, {\bf M}$. 
 Using this fact, our action reads
 \bea
{\cal S} &=& {\cal N}\,\int\,d^4 x\,\left(\partial \pi\right)^2\Big[1+c  \,\left[\Pi\right]+\frac{c^2}{2}\left( 
\left[\Pi\right]^2-\left[\Pi^2\right]\right)
+\frac{c^3}{6}\left(
\left[\Pi\right]^3-3\left[\Pi \right]\left[\Pi^2 \right]+2 \left[\Pi^3 \right]
 \right)
\Big]\,,
\label{actg2}
\eea plus a total derivative. See
also \cite{Tasinato:2013wna} for other uses of a determinantal approach for studying conformal Galileons.
  Action \eqref{actg2} contains a combination of each
Galileon Lagrangian (quadratic, cubic, etc), but with fixed coefficients depending on
powers of the parameter $c$.  We do not include the tadpole contribution. 
  It is straightforward to prove two key properties of Galileon
actions: the corresponding equations of motion (EOMs) are second order, and the theory
enjoys a coordinate-dependent Galileon symmetry \be\label{gal-sym1} \delta
\pi\,=\,w_\mu\,x^\mu+s \ee for constant quantities $w_\mu$, $s$, which is
a symmetry of the action up to boundary terms. 

  In order to get a Galileonic system  with the
preferred coefficients -- say $d_i$ -- in front of each Galileon Lagrangian, we can sum
three copies of action \eqref{actg2} -- each one depending on a parameter $c_i$, with
$i=1, 2, 3$.

Explicitly, we  write 
\be
{\cal S}_{tot}\,=\,\sum_{i=1}^3 S_i\,=\,{{\cal N}}\,\sum_{i=1}^3\,\int \,d^4 x\,\left( \partial \pi \right)^2\,\det{\left[ {\bf 1}+c_i \bf \Pi\right]}\,,
\ee
where each Galileon Lagrangian $S_i$ is characterized by an a priori different constant $c_i$. 
Expanding the determinant in this expression, we find
\bea\label{stotdef1}
{\cal S}_{tot}&=& {\cal N}\,\int\,d^4 x\,\left(\partial \pi\right)^2\Big[d_2+d_3 \,\left[\Pi\right]+\frac{d_4}{2}\left( 
\left[\Pi\right]^2-\left[\Pi^2\right]\right)
+\frac{d_5}{6}\left(
\left[\Pi\right]^3-3\left[\Pi \right]\left[\Pi^2 \right]+2 \left[\Pi^3 \right]
 \right)
\Big]\,,
\eea
with
\bea
d_2&=&3\,{\cal N}
\,,
\\
d_3&=&{\cal N}\left( c_1  + c_2+ c_3 \right)
\,,
\\
d_4&=&{\cal N}\left( 
 c_1^2  +
  c_2^2+
   c_3^2
 \right)
\,, \\
d_5&=&
{\cal N}\left( 
 c_1^3  + c_2^3+ c_3^3
 \right)
\,.\eea
The previous system of  algebraic equations can be solved for $c_i$ in terms of $d_i$ by means of Newton identities.

 \subsection{The duality}\label{gal-dua1}
 The structure of Galileon Lagrangians is  invariant under a {\it duality}, a
 field transformation which connects Galileon theories of different orders (each order
 defined in terms of the number of powers of second derivatives on the scalar field).  The
 properties and consequences of Galileon duality have been introduced in \cite{deRham:2013hsa} and studied at
 length in various works: see e.g. \cite{deRham:2014lqa,Creminelli:2014zxa,Kampf:2014rka,Scargill:2014wya,Noller:2015eda,Noller:2015rea,Baratella:2015yya,Creminelli:2013ygt}.   The duality among different Galileon Lagrangians can  be an important tool to shed
 light on the non-perturbative structure of Galileons, for example to understand  physical 
 consequences of superluminality, and its  connections with screening mechanisms;
 see e.g. \cite{Joyce:2014kja} for a review on these topics.    In this subsection, we briefly review this subject at a formal level, to  demonstrate that 
 a determinantal approach   makes  more   
 manifest the   duality of Galileon actions.
 
  Galileon duality in flat space is based on a field dependent  map among two sets of coordinates, $x^\mu$ and $\tilde x^\mu$
   \footnote{Throughout 
  this work, we 
   use the symbol
  $\Rightarrow$ for denoting the duality transformation.}
 \be \label{dua-def}
x^\mu\Rightarrow \tilde x^\mu\,=\, x^\mu+\frac{1}{\Lambda_S^{3}}\partial^\mu \pi\,,
\ee
where  $\Lambda_S$ is a parameter with dimension of an energy scale, introduced for dimensional reasons. For simplicity, we choose units for  which 
\be\Lambda_S\,=\,1\,\ee
since here (and in what follows) we are more interested to exhibit symmetries and dualities, rather than discussing their physical consequences (strong
coupling scales, etc).

By taking  the differential of eq. \eqref{dua-def}
\bea
d x^\mu&\Rightarrow&d \tilde x^\mu \,=\,\left( \delta_\nu^\mu+\Pi_\nu^\mu \right)\,d x^\nu\,,
\eea
we get
the Jacobian $J^\mu_\nu$ for this transformation 
\be
J^\mu_\nu\,=\, {\color{black}\frac{d \tilde x^\mu}{d x^\nu} } = \,\left( \delta_\nu^\mu+\Pi_\nu^\mu \right) \,.\ee
We assume that such coordinate change is invertible, in the sense that a second  scalar field $\tilde \pi$ exists, 
which sends $ \tilde x^\mu$ back to $x^\mu$ (see \cite{deRham:2014lqa} for details)
\be \label{dua-def2}
\tilde x^\mu\Rightarrow  x^\mu\,=\, \tilde x^\mu-\tilde \partial^\mu \tilde \pi\,.
\ee
We call $\tilde \pi$ the {\it dual field} of $\pi$. 
 
 The requirement of invertibility of this transformation means       that applying the transformations \eqref{dua-def} and \eqref{dua-def2} 
  in succession 
   we go back to the original coordinates:  
the duality is defined as a  map which sends  {the derivative} of the scalar $\pi$ to the derivative of $\tilde \pi$
\bea
\partial^\mu \pi&\Rightarrow&  \tilde \partial^\mu  \tilde \pi\,=\,  \partial^\mu \pi\,.
\eea
So   the {\it derivative of $\pi$} (and not $\pi$)  is a scalar under the duality transformation. On the other hand,  $\pi$ and its dual transform as
\bea
\label{dua1r}
\pi(x)&\Rightarrow&\tilde \pi(\tilde x)\,=\,\pi(x)+\frac12\left( \partial \pi(x)\right)^2\,,
\\
\label{dua1l}
\tilde \pi(\tilde x)&\Rightarrow& \pi( x)\,=\,\tilde \pi( \tilde x)-\frac12\left( \tilde \partial \tilde \pi( \tilde x)\right)^2
\,.
\eea

The second derivative
of $\pi$ transforms non-trivially (as a `{covariant} vector') under duality: using a matrix notation, ${\bf \Pi}\equiv \Pi_\mu^\nu$ we can write
\bea
\label{coot2}
{\bf  \Pi}&\Rightarrow& {\bf \tilde \Pi}\,=\, \left[ {\bf 1}+{\bf \Pi}\right]^{-1}
\,{\bf \Pi}\,.
\eea
where the $\left[\dots\right]^{-1}$ denotes the inverse of a matrix. 
Collecting these results, it is straightforward to determine how the Galileon  system \eqref{actg1} changes under
the action of duality:
\bea {\cal S}\,=\,{\cal N}\, \int d^4 x\,\left( \partial \pi \right)^2\,\det{\left[ {\bf 1}+ c\,{\bf
      \Pi}\right]}&\Rightarrow& {\cal N}\,\int d^4 \tilde x\,\left( \tilde \partial \tilde \pi
\right)^2\,\det{\left[ {\bf 1}+ c\, \tilde {\bf \Pi}\right]}
\,,
\nonumber
\\
&=&{\cal N}\,
 \int d^4 x\,\det\left[ {\bf 1}+\bf \Pi\right]\,\left( \partial \pi
\right)^2\,\det{\left[ {\bf 1}+ c\,\left[ {\bf 1}+\bf \Pi\right]^{-1}\,\bf \Pi\right]}
\,,
\nonumber
\\
&=& 
{\cal N}\,
\int d^4 x\,\left( \partial \pi \right)^2\,\det{\left[\left({\bf 1}+\bf \Pi
    \right)\left( {\bf 1}+ c\,\left[ {\bf 1}+\bf \Pi\right]^{-1}\,\bf \Pi\right)\right]}
\,,
\nonumber
\\
&=& {\cal N}\,
\int d^4 x\,\left( \partial \pi \right)^2\,\det{ \left[ {\bf 1}+ \left(c+1\right)\,\bf
    \Pi\right]}\,=\, \cal \tilde S 
    \,. \label{eq:dual21}
    \eea 
  
    The structure of this determinantal action remains the same, {\bf but}
the coefficient in front of the matrix $\bf \Pi$  within the determinant argument passes from the value $c$ to $(c+1)$. This is the  only change induced
by  applying the duality.

 We can then combine different dual actions ${\cal \tilde  S}$, to form an arbitrary combination of Galileon
Lagrangians with arbitrary coefficients (as done around eq. \eqref{stotdef1}). Comparing the results before and
after applying the duality transformation, we  find that 
the overall coefficients in front of each  Galileon Lagrangian are mapped to
their  `dual' values 
 \bea \tilde d_2&=&3\,{\cal N}\,,
\\
\tilde d_3&=&{\cal N}\left( c_1 + c_2+ c_3+3 \right)\,,
\\
\tilde d_4&=&{\cal N}\left( \left(c_1+1\right)^2 + \left(c_2+1\right)^2+
  \left(c_3+1\right)^2 \right)
\,,\\
\tilde d_5&=& {\cal N}\left( \left(c_1+1\right)^3 + \left( c_2+1\right)^3+
  \left(c_3+1\right)^3 \right)\,, \eea 
 where the $d_i$ are the quantities introduced in eq. \eqref{stotdef1} and following.
  Hence we have the relations
   \bea \tilde d_2&=& d_2
\,,\\
\tilde d_3&=& d_2+d_3
\,,\\
\tilde d_4&=& d_2+2 d_3+d_4
\,,\\
\tilde d_5&=& d_2+3 d_3+3 d_4+d_5 \,.\eea We checked that these results are in  agreement
with \cite{deRham:2013hsa}  \footnote{The dictionary between the notation of \cite{deRham:2013hsa} and
  ours is as follows: {{$d_2 = - 6 b_2, d_3 = 3 b_3, d_4 = -4 b_4, d_5 = 15 b_5, 
      \tilde d_2 = - 6 p_2, \tilde d_3 = -3 p_3, \tilde d_4 = -4 p_4, \tilde d_5 = -15 p_5$,
      where we renamed their $c_i$'s to $b_i$'s. These relations take care of differences in the
definitions of the Galileon Lagrangians -- total derivatives and global coefficients -- as well as of 
a sign difference in the definition of the dual field. 
}}}. {Note in passing  that \eqref{eq:dual21} provides a simple expression for the dual of  free fields, once we select $c_i\,=\,0$.}

\section{Symmetries and dualities for DBI Galileons}\label{sec-dbigals}

\subsection{Some motivations}

One motivation for introducing Galileons is to find a `ghost-free' version of a system describing the 
physics of  the  DGP brane-world \cite{Dvali:2000hr}.  It is then natural to ask whether  Galileon actions
  have 
a geometrical description in terms of   a brane embedded in higher dimensional space. This was achieved in \cite{deRham:2010eu}, and
generalised in  \cite{Goon:2011uw,Goon:2011qf}, introducing a class of theories 
called DBI Galileons. They enjoy symmetries generalising Galileon transformations (in absence of dynamical  gravity).
 In this work  we show that the same approach can be used to  find a relation between DBI  Galileons and   
a subclass of beyond Horndeski and  EST theories.  
 First of all, however, to pave the way we need to reconsider DBI Galileons from a perspective which is  slightly different
from the one of \cite{deRham:2010eu}.

\smallskip

The approach of \cite{deRham:2010eu} starts from the fact that
brane    actions   can be built  by means of gravitational Lovelock and Gibbons-Hawking terms, which describe derivative self-interactions for  a scalar  controlling the position of the probe brane in the
 higher dimensional bulk. The resulting scalar actions are automatically ghost free, since they are built in terms of specific   
 combinations (DBI, Lovelock, Gibbons-Hawking) of the brane induced  metric,  ensuring  that the
 scalar  equations of motion are at most second order. In addition, this  scalar action enjoys symmetries inherited from  
  isometries of the higher dimensional space. These symmetries, associated with the properties of the 
 extra dimensional geometry, can  reduce to Galileon symmetries
 in appropriate, `small first derivative' limits.
 
 \smallskip
 
 On the other hand, recently it has been realised that degenerate scalar-tensor theories  exist, which although characterized 
 by higher order EOMs, are nevertheless consistent  thanks to the existence of primary constraints. These are the theories
 of beyond Horndeski, or more in general EST/DHOST \cite{Langlois:2015cwa,Crisostomi:2016czh,Achour:2016rkg}. It is natural to ask whether 
 these theories admit some sort of geometrical interpretation.
  This is one of the purposes of our work, and we start in this Section to examine scalar  theories
 which will be   related to degenerate scalar-tensor theories, once  coupled to dynamical  gravity.  
In particular, in this Section we build on the results of \cite{deRham:2010eu,Goon:2011uw,Goon:2011qf},
 but we discuss   convenient, determinantal expressions for  
 DBI Galileons. This  allows us to express
 such actions in a more  compact form, and to exhibit symmetries and dualities among them.  
 Our method for expressing
 the DBI Galileon system does {\it not  directly rely} on using Lovelock or Gibbons-Hawking combinations: on the other hand,  
  it  provides  consistent theories   in absence of dynamical gravity. Our approach 
       will then  be used in Section \ref{sec-est} to make  manifest the connection between DBI Galileons and
  beyond
 Horndeski/EST  theories, once the system is minimally coupled with dynamical gravity. 
 
  We start discussing 
 Poincar\'e  DBI Galileons  -- so called since they correspond to scalar actions associated with  a brane embedded
 in a 5d bulk with Poincar\'e symmetry -- to then continue analysing AdS DBI Galileons. 

\subsection{Poincar\'e  DBI Galileons }\label{sec-poidbi}

Poincar\'e 
DBI Galileons \cite{deRham:2010eu,Goon:2010xh,Hinterbichler:2010xn} are scalar theories with second
order EOMs, enjoying a symmetry that generalises
Galileon invariance. We  call such symmetry Poincar\'e induced symmetry, being
inherited from a global 
Poincar\'e symmetry in five dimensions.
  Namely, their action and the symmetry they satisfy  are   inherited from  a five dimensional description in terms
of a probe brane in a 5d flat  space.  
 The system can   
 also be described 
  using a 
    determinantal
 approach,  which generalises the  one applied in the previous section to standard Galileons.
 
 \smallskip
 
 The action we are interested in   
 reads
\be \label{sdbi2}
 S\,=\,{\cal N}\,\int d^4 x\,\frac{1}{\gamma}\det{\left[ \,\delta_\mu^\nu+ {c}\,{\gamma}
\left(\Pi_\mu^\nu-\gamma^2\,
\partial_\mu \pi\,\partial^\lambda \pi\,\Pi_\lambda^\nu\,\right)
\right]}\,,
\ee
with
\be \label{gadef1}
\gamma\,=\,\frac{1}{\sqrt{\kappa_0^2+X}}\,.
\ee
 ${\cal N}$,  $c$, $\kappa_0$ are constants, and
\be
X\,=\,(\partial \pi)^2\,.
\ee 
 The motivation for considering  this action will be clearer in what follows. 
  When expanding the determinant, one finds a sum of
four actions, weighted by different powers of the constant parameter $c$,  from zero to three.  A direct calculation shows that each of them
reproduces the structure of the DBI Galileon Lagrangians presented in
\cite{deRham:2010eu}, once selecting $\kappa_0=1$.  Hence, action \eqref{sdbi2} succinctly contain all DBI Galileons in flat space. 
More explicitly, 
using eq. \eqref{det-for}, 
we expand the determinant and get
\bea
S&=&{\cal N}\,\int d^4 x\,\frac{1}{\gamma} \Big(1+ c \gamma \left( [\Pi]-\gamma^2 [\Phi]\right) 
+\frac{c^2\,\gamma^2}{2}\left( [\Pi]^2-[\Pi^2] +2 \gamma^2 [\Phi^2]-2 \gamma^2 [\Phi] [\Pi] \right)
\nonumber\\
&&+\frac{c^3\,\gamma^3}{6}\left[
[\Pi]^3+2 [\Pi^3]-3 [\Pi^2]  [\Pi] + 3 \gamma^2 \left(2 [\Pi] [\Phi^2]-2 [\Phi^3] - [\Phi] [\Pi]^2+
 [\Phi] [\Pi^2] \right)
\right]
\Big)\,,
\eea
where we use the notation $\left[ \Phi^n\right]\,=\,{\text{ tr}}\left(\partial  \pi  \,\Pi^{n}
\, \partial \pi \right)$.  The coefficient of each power of $c$ corresponds to one of the Lagrangians for DBI Galileons, when  choosing 
$\kappa_0=1$. 
 A linear combination of these actions
is able to provide DBI Galileons with arbitrary coefficients (as described in the previous
section for the standard Galileon case).   Working within a determinantal  approach  allows us 
to make more transparent dualities and symmetries for DBI Galileons. 

 Notice that
   we are discussing a slight generalisation
of the actions of \cite{deRham:2010eu}, which includes a free constant parameter
$\kappa_0$ in the definition of the $\gamma$ factor~\footnote{The generalisation we
  consider is a particular case of the actions discussed in \cite{Goon:2011uw,Goon:2011qf}. It is simple to see
  that this additional free parameter can be obtained from the standard DBI Galileon case,
  by a rescaling $\pi \to \pi/\kappa_0$.}.  We do so because once coupled with gravity, 
appropriate  
    choices of this parameter $\kappa_0$
   make manifest the connection between these actions and degenerate scalar-tensor theories. This  will be
   discussed  in Section \ref{sec-est}. 
   

Action \eqref{sdbi2} leads to second order equations of motion for the scalar field,
thanks to the properties of the determinant. Additionally, this action is invariant (up to boundary terms)
under a  scalar 
symmetry (here $\omega^\mu$ is an arbitrary constant vector)
\be \label{symdbi2}
\delta \pi\,=\,\kappa_0^2\,\omega_\mu\,x^\mu+\pi\,\omega^\mu\,\partial_\mu\,\pi\,,
\ee
 and  under  a duality, as we  discuss in the next two subsections.

\subsubsection{Geometrical interpretation, and underlying symmetry}\label{sym-poi}
As we mentioned, the scalar Lagrangian \eqref{sdbi2} is invariant under  the scalar transformation \eqref{symdbi2},
 up
to boundary terms 
   (an additional, shift symmetry $ \pi\to  \pi+$const is also satisfied, but we do not consider it  here). This can 
   be proved by a 
%
 direct computation.  In the  limit of small field derivatives, this transformation reduces to Galileon symmetry, at least when
$\kappa_0\neq 0$ (more on this later). 

Alternatively, this symmetry can be understood `geometrically' in terms of an action for a  probe brane embedded in a higher
dimensional bulk, 
 using  arguments similar to   \cite{deRham:2010eu} --  further developed in \cite{Goon:2011uw,Goon:2011qf} -- that
we briefly review  here, and  accommodate to our discussion.  

The transformation   \eqref{symdbi2} is associated with a symmetry  for a probe brane in 5d flat space,   inherited from a global 
 isometry in five dimensions. In particular, eq. \eqref{symdbi2} is associated with  boosts in five dimensions.
 To see this fact more explicitly,
   we consider   a 5d  bulk with  flat metric 
\be g^{(5)}_{MN}\,dX^M d X^N\,=\,\kappa_0^2\,\eta_{\mu\nu} \,
d X^\mu d X^\nu+ d y^2 \,,\ee 
{\color{black}where $X^5 = y$}. We introduce a constant parameter $\kappa_0^2$ in front of the 4d slices in the 5d metric.  Still, the 5d metric
is flat, and have the same number of isometries of Minkowski space.
As we will discuss in Section \ref{sec-dbiextr}, 
 the parameter $\kappa_0$ is associated with the `maximal speed'
allowed by causality for motion along the extra dimension. 

\begin{figure}[hth!]
\centering
\scalebox{0.4}{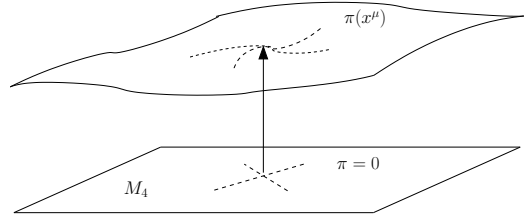}
\caption{Brane geometry with respect to a bulk foliation $y=const$. }\label{BraneF}
\end{figure}

A 4d brane embedded in the 5d bulk is characterised by a brane embedding, $X^M(x^\mu)$, which maps 
the four brane dimensions into the five  bulk  dimensions.  We foliate the bulk in terms of slices $y=$const; the brane
embedding is chosen as
\bea \label{bre1}
X^\mu&=& x^\mu\,,
\nonumber
\\
y&=& \pi(x)\,,
\label{gbe1}
\eea
and fixes the gauge associated with the freedom to reparameterise the foliation.  
    The scalar field $\pi$ is a modulus which geometrically corresponds
to the position of the brane along the fifth bulk coordinate. See Figure \ref{BraneF}.

The brane induced geometry can be deduced from the information we provided.
  The induced brane metric is
\be \label{pdbi-g}
g_{\mu\nu}\,=\, \frac{\partial X^M}{\partial x^\mu}\frac{\partial X^N}{\partial x^\nu} g^{(5)}_{MN}\, =  \,\kappa_0^2 \eta_{\mu\nu}+\partial_\mu \pi \partial_\nu \pi\,.
\ee
 The {matrix inverse of the induced brane metric is} 
\be
g^{\mu\nu}\,=\,\frac{1}{\kappa_0^2}\left( \eta^{\mu\nu}-\gamma^2\partial^\mu \pi \partial^\nu \pi\right),
\ee
where recall that {$\gamma = (\kappa_0^2 + X)^{-1/2} $}.
It satisfies the relation
\be
g^{\mu\,\alpha}\,
 g_{\alpha\nu}\,=\,\delta^\mu_{\nu}\,.
 \ee
 The square root of the metric determinant is
 \be
 \sqrt{-g}\,=\,\frac{\kappa_0^3}{\gamma}
 \,.\ee
Another tensorial quantity of interest is proportional to the `brane extrinsic curvature', a tensor defining intrinsic properties of the brane geometry which can be computed
using standard definitions \cite{Wald:1984rg}. In this case, it results
\be
K_{\mu\nu}\,=\,-\gamma\,\kappa_0\,\Pi_{\mu\nu}\,.
\ee
An  interesting property of the quantities $g_{\mu\nu}$, $K_{\mu\nu}$ is that they transform as tensors with respect
to the transformation \eqref{symdbi2}, which we rewrite here
\be \label{symdbi2a}
\delta \pi\,=\,\kappa_0^2\,\omega_\mu\,x^\mu+\pi\,\omega^\mu\,\partial_\mu\,\pi
\,.
\ee
Namely, when applying the
scalar transformation \eqref{symdbi2a},   these quantities  transform as 
\be
g_{\mu\nu}\,\to\,\xi^\alpha\,\partial_\alpha g_{\mu\nu}+\partial_\mu \xi^\alpha\,g_{\alpha \nu}
+\partial_\nu \xi^\alpha\,g_{\alpha \mu}
\,.
\ee
(and analogously for $K_{\mu\nu}$)  
with
\be
\xi^\alpha\,=\,\pi\,\omega^\alpha\,,
\ee
and $\omega^\alpha$ being  the constant vector of  eq. \eqref{symdbi2a}.  

\smallskip

The scalar symmetry \eqref{symdbi2a} can be interpreted as geometrically associated with isometries of the embedding   5d
geometry: this viewpoint has been developed in a comprehensive way in   \cite{Goon:2011uw,Goon:2011qf}. 
  Suppose we have a bulk isometry associated with a Killing vector ${\mathcal
  V}^A$: the probe brane action  should enjoy this symmetry as well, in the form of a  
  symmetry 
  transformation for the scalar field $\pi$.
   On the other hand, we
  have to take into account  that by choosing the embedding \eqref{gbe1} we are selecting a specific
  gauge, associated with our freedom of choosing the brane coordinates. As explained in  \cite{Goon:2011uw,Goon:2011qf}, 
  we need to include a `compensating' gauge transformation to the field $\pi$, for ensuring that
  the brane action is invariant under bulk isometry. In total, the scalar transformation inherited from the bulk isometry, which is a
  symmetry of the brane action, reads
 \be \delta \pi\,=\,\mathcal V^5(x,\pi)-\kappa_0\,\mathcal V^\mu \partial_\mu\pi\,, \ee 
  \smallskip
where the second term in the right hand side is associated with the compensating  gauge transformation. Applying these
arguments to our case, we find the scalar symmetry \eqref{symdbi2a}.

\smallskip

A consequence of the   symmetry is that any action which is a scalar built in terms of the
tensors $g_{\mu\nu}$, $K_{\mu\nu}$ is invariant under the transformation \eqref{symdbi2}.
Among the symmetry preserving actions, we find the determinantal action
\eqref{sdbi2} we considered in the previous subsection, which can be expressed as

\bea 
S&=&
 \frac{{\cal N}}{\kappa_0^3}\,\int d^4 x\, \sqrt{-g}\,\det{\left[ \delta^\mu_\nu-\kappa_0\,c\,g^{\mu\,\alpha}
    K_{\alpha\,\nu} \right]}\,,
\label{sdbi2aaa}
\\
\label{sdbi2a}
&=& {\cal N} \,\int d^4 x\,\frac{1}{\gamma}\det{\left[ \,\delta^\mu_\nu+
    {c}\,{\gamma} \left(\Pi^\mu_\nu-\gamma^2\,
      \partial^\mu \pi\,\partial_\lambda \pi\,\Pi^\lambda_\nu\,\right) \right]}\,.
\eea 
Action \eqref{sdbi2a} coincides with eq. \eqref{sdbi2}. 
Hence it is a scalar action in flat 4d space,  built with the tensors
$g_{\mu\nu}$, $K_{\mu\nu}$; and it is invariant under symmetry \eqref{symdbi2} up to boundary
terms: calling ${\bf K }\, =\,g^{\mu\,\alpha}
    K_{\alpha\,\nu}$, expanding the determinant we find combinations   of traces tr$ {\bf K}$,  tr$ {\bf K^2}$, $\dots$, 
    and their powers, 
 Remarkably, as stated previously, we get all the four
combinations that correspond to DBI Galileons. The determinantal action, additionally, is
useful for exhibiting a duality for DBI galileons as we discuss in  the next subsection.

 It
is important to emphasize again  that our derivation of DBI Galileons is different with respect
to the approach of \cite{deRham:2010eu}.
 In that case, DBI Galileons are obtained starting
from {combinations of} curvature invariants that automatically ensure that the EOMs
for the fields involved are second order (Lovelock and Gibbons-Hawking terms); this
remains true when gravity is made  dynamical (i.e. the flat metric $\eta_{\mu\nu}$ on 4d slices $y=const$
is promoted to a dynamical metric $q_{\mu\nu}$), and allows one to obtain covariantized versions of Galileons \cite{deRham:2010eu}. 

In 
our approach, working with the determinantal action \eqref{sdbi2a}, we are not ensured that
the EOMs remain of second order, once the theory is minimally coupled with gravity by making gravity dynamical.
 Indeed, when coupled with gravity, the system is characterised by higher order EOMs. On
 the other hand, as we will see in Section \ref{sec-est},  when $\kappa_0\to 0$ 
  a  primary  constraint arises, which prevents the propagation of an additional degree of freedom associated with an Ostrogradsky instability:
  the resulting theory belongs to the class of beyond Horndeski theories of gravity, or more generally to EST.

\subsubsection{The duality}\label{dua-poi}


Action \eqref{sdbi2a} satisfies a duality which generalises the Galileon duality we reviewed in Section \ref{gal-dua1}.
 Our determinantal approach to DBI Galileon makes this duality manifest, and as far as we are aware we are the
 first to  discuss in this particular  way a duality for Poincar\'e  DBI galileons. 
  (See also \cite{Creminelli:2014zxa,Creminelli:2013ygt}.) 

Our arguments 
here are very similar in spirit to the discussion of duality for standard Galileons, as developed in Section \ref{gal-dua1}.
 The determinantal
action \eqref{sdbi2}  can be expressed 
in several equivalent ways
\bea
S&=&
{\cal N}\,\int d^4 x\,\frac{1}{\gamma}\det{\left[ \,\delta_\mu^\nu+ {c}\,{\gamma}
\left(\Pi_\mu^\nu-\gamma^2\,
\partial_\mu \pi\,\partial^\lambda \pi\,\Pi_\lambda^\nu\,\right)
\right]}\,,
\\ &=&
{\cal N}\,\int d^4 x\,\frac{1}{\gamma}\det{\left[ \,\delta_\mu^\nu+ {c}\,
\partial_\mu \left(\gamma \partial^\nu \pi\right)
\right]}\,,
\\
&=& \frac{{\cal N}}{\kappa_0^3}\,
\int d^4 x\, \sqrt{-g}\,\det{\left[
\delta^\mu_\nu-\kappa_0\,c\,g^{\mu \sigma} \,K_{\sigma\,\nu}
\right]}\,.
%
 \label{ac-dbi-sec}
\eea
where in the last line we write  the action in a `geometric form', exploiting
 the concept of induced metric and extrinsic curvature on a probe brane, as discussed in the previous
 Subsection. Recall that $\gamma^{-1}\,=\,\sqrt{\kappa_0^2+X}$. 
 
We introduce  a field dependent coordinate map which is at the basis of the duality
\be\label{trdbig0}
x^\mu\,\,\Rightarrow\,\,
\tilde x^\mu\,=\,x^\mu+\gamma\,\partial^\mu \pi\,,
\ee
(where, as for the case of Galileons, Section \ref{gal-dua1},   we choose units which set to one the parameter  $\Lambda_S$ which is 
 needed for dimensional reasons). We demand that there exists a `dual' scalar $\tilde \pi$, which maps through duality the tilde
 coordinates $\tilde x^\mu$ back to the original coordinates $x^\mu$:
 \be\label{trdbig20}
\tilde x^\mu\,\,\Rightarrow\,\,
 x^\mu\,=\,\tilde x^\mu-\tilde \gamma\,\tilde \partial^\mu \tilde \pi\,.
\ee
 Comparing eqs \eqref{trdbig0} and \eqref{trdbig20}, we find 
\be \label{cod1aa}
\tilde \gamma\,\tilde \partial^\mu \tilde \pi\,=\,\gamma\,\partial^\mu \pi
\,.
\ee
The simplest way to satisfy the previous condition \eqref{cod1aa} is to impose
\be \label{cod1ab0}
\tilde \partial^\mu \tilde \pi\,=\,\,\partial^\mu \pi\,,
\ee
that is, the duality maps the derivative to the scalar $\pi$ to the derivative of the dual $\tilde \pi$. 

The Jacobian associated with map  \eqref{trdbig0} is

\bea
J^\mu_\nu&\equiv&\frac{d \tilde x^\mu}{d x^\nu}   \,=\, \delta^\mu_\nu+\gamma\left(\delta^\mu_\rho-\gamma^2 
\partial^\mu\, \pi \partial_\rho \pi\right)\,\Pi^\rho_\nu
\,,
\\
&=&\delta^\mu_\nu
-\kappa_0\,g^{\mu\,\alpha} K_{\alpha\,\nu}\,.
\eea 
Eq \eqref{cod1ab0} implies (indexes are raised/lowered with flat Minkowski metric)
 
 \be
 \tilde \partial_\mu \tilde \pi\, d \tilde x^\mu =\,\,\partial_\mu \pi\,J^\mu_\nu\,d x^\nu\,=\,\partial_\mu \pi\, d x^\mu+\partial_\mu \pi\, \partial_\nu
 \left( \gamma \partial^\mu \pi\right)d x^\nu\,.
 \ee

 Integrating, we find a non local  relation among the field $\pi$ and its dual
  \be
  \pi(x)\,\Rightarrow\, \tilde \pi(\tilde x)\,=\, \pi(x)+ \int^x\,d \bar x^\lambda\,\partial_\lambda\left(\gamma\partial^\rho \pi\right)\,\partial_\rho \pi
 \,.\ee
The second derivative of the scalar, on the other hand, transforms under duality by means of the inverse Jacobian:
\be
\partial_\mu \left( \gamma \partial^\nu \pi \right) \hskip1cm \Rightarrow \hskip1cm
\tilde \partial_\mu \left(  \tilde \gamma \tilde  \partial^\nu \tilde \pi \right)\,=\,\tilde \partial_\mu \left(\gamma \partial^\nu \pi \right)
\,=\,(J^{-1})_{\mu}^\lambda\, \partial_\lambda \left(\gamma \partial^\nu \pi \right)\,.
\ee
 
 Notice that, using the definitions of induced brane geometrical quantities
  \bea
 g_{\mu\nu}&=& \kappa_0^2 \eta_{\mu\nu}+\partial_\mu \pi \partial_\nu \pi
 \\
\kappa_0\,g^{\mu\alpha}\, K_{\alpha\nu}&=&- \partial^\mu \left( \gamma \partial_\nu \pi\right)
\,.
 \eea
The results so far imply that
 \bea
 d^4 x&\Rightarrow&  d^4 \tilde{x}\,=\,\left(\det{J}\right)\, d^4 x \,,\\
 \sqrt{-g}&\Rightarrow&   \sqrt{-\tilde g} \,=\, \sqrt{-g}\,,\label{dua-det} \\
 g^{\mu\alpha }K_{\alpha \nu}&\Rightarrow&   \tilde g^{\mu\alpha } \tilde{K}_{\alpha \nu}\,=\,\left(J^{-1}\right)^\mu_{\rho}\, g^{\rho \alpha}\,K_{\alpha \nu}
 \,.
 \eea
 These ingredients are sufficient for finding how our original action transforms under duality (here $K_\nu^\mu\,=\,g^{\mu\alpha }K_{\alpha \nu}$)
 \bea
{\frac{\cal N}{\kappa_0^3}}
\int d^4  x\, \sqrt{- g}\,\det{\left[
\delta^\mu_\nu-\kappa_0\,c\,  K_{\,\nu}^\mu
\right]}
&\Rightarrow&
{\frac{\cal N}{\kappa_0^3}}
\int d^4 \tilde x\, \sqrt{-\tilde g}\,\det{\left[
\delta^\mu_\nu-\kappa_0\,c\, \tilde K_{\,\nu}^\mu
\right]}\nonumber
\\
%
%
&=&
{\frac{\cal N}{\kappa_0^3}}
\int d^4  x\,\det{J}\,\sqrt{- g}\,\det{\left[
\delta^\mu_\nu-\kappa_0\,c\, \left( J^{-1}\right)^{\mu}_\beta \, K^\beta_{\,\nu}
\right]} \nonumber
\\
&=&
{\frac{\cal N}{\kappa_0^3}}
\int d^4  x\, \sqrt{- g}\,\det{\left[
\delta^\mu_\nu-\kappa_0\,(c+1)\,   K^\mu_{\,\nu}
\right]}\,.
\eea
Hence  the structure of the action remains the same, and the only change is a shift  in the constant 
$c \Rightarrow c+1$ which appears within the determinant. This result
generalises the standard
Galileon duality that we reviewed in Section \ref{gal-dua1}.  

   \subsection{DBI Galileons in a maximally symmetric extra dimensional space}\label{sec-adsdbi}
  
  
  Some of the results we discussed in the previous subsections can be generalised to a set
  of scalar actions associated with branes probing curved  5d space times, as for example
  AdS or dS spaces, which maintain a four dimensional flat slicing.  
   The new feature introduced by such versions of DBI
  Galileons is an explicit dependence of the action on the field $\pi$ (and not only on
  its derivatives)  and a generalisation of the symmetries reviewed earlier. These actions fall in the class of conformal Galileons, in the nomenclature 
  of 
\cite{Goon:2011uw,Goon:2011qf}.
 
In order to discuss these theories, we use the convenient geometrical  approach introduced in \cite{Goon:2011uw,Goon:2011qf}. We consider for definiteness a curved 5d space time with warped metric 
\be\label{5dma}
ds^2_{(5)}\,=\, \kappa_0^2\,
f^2(y)\,\eta_{\mu\nu}\,d X^\mu d X^\nu+ d y^2
. 
\ee
We examine  the same brane  {\color{black}configuration} as in the previous subsection
\bea
X^\mu&=& x^\mu,
\\
y&=& \pi(x^\mu).
\eea
The associated brane induced metric results
\be \label{fmet}
g_{\mu\nu}\,=\,f^2(\pi)\,\kappa_0^2\,\eta_{\mu\nu}+\partial_\mu \pi \partial_\nu \pi\, ,
\ee
with inverse
\be \label{fumet}
g^{\mu\nu}\,=\,\frac{1}{f^2\,\kappa_0^2}\,\left(\eta^{\mu\nu}-\frac{\partial^\mu \pi\, \partial^\nu \pi}{\kappa_0^2\,f^2+X} \right).
\ee
 The brane extrinsic curvature is
 \be\label{fextcur}
 K_{\mu\nu}\,=\,-\frac{\kappa_0\,f}{\sqrt{\kappa_0^2 f^2+X}}\,\left( \partial_\mu \partial_\nu \pi
 {\color{black}-}\frac{2 f'}{f}\,\partial_\mu \pi \partial_\nu \pi-\kappa_0^2\,f\,f'\,\eta_{\mu\nu}\right)\, .
 \ee
 {We construct an action with the very same geometric structure as the one of the previous subsection (see eq. \eqref{sdbi2a},} 
\bea \label{lagadsdbi}
{\cal S}\,=\, \frac{\cal N}{\kappa_0^3}\int d^4 x\, \sqrt{-g} 
\det{\left[
\delta_\mu^\nu-\,c\,\kappa_0\, g^{\mu \alpha}\,K_{\alpha \nu}
\right]}\, ,
\eea
but with  the new induced metric and brane extrinsic curvature  given in eqs \eqref{fmet}, \eqref{fextcur}.
As explained in Section \ref{sec-poidbi}, and in detail in  \cite{Goon:2011uw,Goon:2011qf},  being a scalar built in terms of the tensors $g_{\mu\nu}$, $K_{\mu \nu}$, this action is invariant
 under any scalar symmetry associated with the isometries 
of the 5d space under consideration. 
 
  \bigskip

 As a concrete example, that turns to be relevant  for what comes next,  we consider a probe Minkowski brane embedded in a 5d AdS bulk. This embedding is described by the warp factor
 \be \label{chof}
 f(\pi)\,=\,e^{- \frac{ \pi}{\ell}}\, ,
 \ee
 in eq. \eqref{5dma}, 
 with $\ell$ the AdS radius. In the limit $\ell\to \infty$ we recover $f=1$ and the Poincar\'e DBI Galileons
 of  subsection \ref{sec-poidbi}.  
Plugging \eqref{fumet}, \eqref{fextcur} and \eqref{chof} in \eqref{lagadsdbi} we obtain   
 \bea 
 {\cal S}&=& {\cal N}\, \int d^4 x\,\frac{e^{-3 \pi/\ell}}{\gamma}\,
\det \left[\frac{}{}
\delta^\mu_\nu 
+ 
c\,\gamma\, e^{\pi/\ell} \,\left(\eta^{\mu\alpha}-
\gamma^2
{\partial^\mu \pi\, \partial^\alpha \pi} \right)
\,\left( \partial_\alpha \partial_\nu \pi
 {\color{black}+}\frac{2 }{\ell}\,\partial_\alpha \pi \partial_\nu \pi+\frac{ \kappa_0^2}{\ell}\,e^{-2 \pi/\ell}\,\eta_{\alpha\nu}\right)
\right]\,,
\nonumber \\ 
 \label{acaddb1}
\eea 
with 
\be 
\gamma\,=\,\frac{1}{ \sqrt{\left( \kappa_0^2\,e^{-2 \pi/\ell}+X\right) } }.
\ee 
This action is invariant under the following  transformation of the
scalar field, which is a symmetry inherited from the isometries of the AdS bulk
\be \label{scsads1} \delta \pi\,=\,\kappa_0^2 \,w_\mu x^\mu+\partial_\mu \pi\,\left(
  \frac{\ell}{2}\left( e^{2 \pi/\ell}-1\right)\,w^\mu+\frac{\kappa_0^2\; x^2}{2
    \ell}\,w^\mu- \frac{\kappa_0^2}{\ell} \,\left( w x \right)\,x^\mu \right).  \ee When
expanding the determinant, one finds a set of actions which are related to the
AdS DBI Galileons discussed in \cite{deRham:2010eu}\footnote{One finds the Lagrangians for
  AdS DBI Galileons expanding the determinant of a slightly different action, given by
  \bea \label{lagadsdbib} {\cal S}\,=\, \int d^4 x\, \sqrt{-g} \,\left\{ \det{\left[
        \delta^\mu_\nu-\,c \, \kappa_0\,g^{\mu \alpha}\,K_{\alpha \nu} \right]}
    -\frac{6\,c^2 \, \kappa_0^2}{\ell^2}+\frac{3\,c^3 \, \kappa_0^3}{2\,\ell^2}\,K
  \right\}\,.  \eea when setting $\kappa_0=1$. On the other hand, both actions
  \eqref{lagadsdbi} and \eqref{lagadsdbib} are invariant under the scalar symmetry
  \eqref{scsads1} and have second order equations of motion. }. The advantage of such
geometrical approach,  developed in \cite{deRham:2010eu,Goon:2011uw,Goon:2011qf}, is that it makes more manifest the symmetries associated with the
action \eqref{lagadsdbi}.
 For the case of AdS DBI Galileons, however, it is not clear whether a duality exists which
connects actions of different order.  The case of a brane embedded in de Sitter space is discussed
in Appendix \ref{app-des}.
 
 \section{An extreme relativistic limit}\label{sec-dbiextr}
 
 We examine in this Section  a certain limit of the DBI Galileons in flat space, which we dub extreme relativistic,
   which satisfies a symmetry different from Galileon symmetry.
   As
 we will learn in the next Section, 
 the resulting
 theories are particularly interesting when coupled with dynamical gravity,  since  they are related with beyond Horndeski
 and other degenerate scalar-tensor theories. 
 
 When discussing Poincar\'e DBI Galileons, we  considered a five dimensional flat metric, characterised
 by a parameter $\kappa_0$ as
 \be
 d s^2\,=\,\kappa_0^2\,\eta_{\mu\nu} d x^\mu d x^\nu+ d y^2\,.
 \ee
 Physically, the parameter $\kappa_0$ is a `warp factor'  controlling the maximal velocity along the fifth  dimension.  
 The speed of light ${\bf v}_{light}^{y}$ along the  extra dimension $y$ is 
 \be \label{spex}
 {\bf v}_{light}^{y}\,=\,\kappa_0\,,
 \ee
 in units where   on  the four dimensional slices $y\,=\,const$ the speed of light ${\bf v}_{light}^{4d}$ is       ${\bf v}_{light}^{4d}=1$.
  
 The dynamics of the brane in the extra dimension  depends on the value of $\kappa_0$. Recall   that,  for
  Poincar\'e DBI Galileons,  we are dealing with the
 brane action
 \be\label{act-ag1}
  S\,=\,{\cal N}\,\int d^4 x\,\frac{1}{\gamma}\det{\left[ \,\delta_\mu^\nu+ {c}\,{\gamma}
\left(\Pi_\mu^\nu-\gamma^2\,
\partial_\mu \pi\,\partial^\lambda \pi\,\Pi_\lambda^\nu\,\right)
\right]}\,,
\ee
 where $\gamma^{-1}\,=\,\sqrt{\kappa_0^2+X}$. The action  is invariant under the scalar transformation 
 \be\label{sym-ag1}
 \delta \pi\,=\,\kappa_0^2\,\omega_\mu\,x^\mu+\pi\,\omega^\mu\,\partial_\mu\,\pi\,
 .\ee
  We can   distinguish
  three physically distinct cases:

  \begin{itemize}
  \item {\underline{Standard DBI Galileons}: $\kappa_0=1$}.  The  speed of light along the extra dimension is the same as in the four dimensional slices.  
  The scalar action  can be interpreted geometrically in terms of a brane probing an extra dimensional
  space time, as  reviewed in Section \ref{sec-poidbi}.  It enjoys 
   the symmetry \eqref{sym-ag1} for $\kappa_0=1$ which generalises  the Galilean symmetry adding a relativistic correction to it.
  \item {\underline{5d non relativistic limit}: $\kappa_0\to \infty$}. In this limit, the speed of light along the extra dimension
  is infinite, see eq. \eqref{spex}. We expect then that   there is no causal bound on the speed
  along the extra dimensions.  Indeed  the theory that we get corresponds to standard Galileons:
  the first derivatives of the brane modulus  $\pi$ are small, and relativistic  corrections are negligible.  In
  this limit, the factor 
    $\gamma\, \simeq \kappa_0^{-1}\to 0$. To  find a meaningful action, at  the same time  we then select large values for the constants $\cal N$ and $c$ such 
   that  ${\cal N}/\kappa_0\,=\,2\,{\bar {\cal N}}\,=\,const$ and ${c}/\kappa_0\,=\,{\bar {c}}\,=\,const$. 
   In this limit,
     we then find the  following action for the system
   \be
    S_{\kappa_0\to \infty}\,=\,{\bar {\cal N}}\,\int d^4 x\,\left( \partial \pi \right)^2\,\det{\left[ \,\delta_\mu^\nu+ \bar{c}\,
\Pi_\mu^\nu
\right]}
   \ee
plus a total derivative,    which corresponds to  the standard Galileon action of eq. \eqref{actg1}. 
  \item {\underline{5d extreme relativistic limit}: $\kappa_0\to 0$}.   In this limit, the speed of light along the extra dimension
  $y$ vanishes~\footnote{This is reminiscent to what happens in a black hole geometry, where the `speed of light' vanishes at the black hole 
  ergosurface where the coefficient $g_{tt}$
  of the time coordinate vanishes.    It would be interesting to pursue this analogy further and reformulate the $\kappa_0\to0$ limit as approaching a special point on 
   some examples of
   5d geometries.}: this is a peculiar limit, where causality forbids a motion along the extra dimension $y$. The brane action
  results
  \bea\label{actk01a1}
 S_{\kappa_0\to 0}\,=\,{\cal N} \int d^4 x\,  \sqrt{X}  \,
\det{\left[
\delta^\mu_\nu+\,c \frac{1}{ \sqrt{X}}\left(
\,\delta^\mu_\rho-\frac{\,\partial^\mu \pi  \partial_\rho \pi}{X}
\right)
 \,
 \partial^\rho \partial_\nu \pi
\right]}\, .
\eea
In order to have a well defined square root, $X>0$, and this implies that we need to focus brane actions with space like
 scalar first derivatives $\partial \pi$. In this extreme relativistic limit, the action has still a symmetry
 \be
  \delta \pi\,=\,\pi\,\omega^\mu\,\partial_\mu\,\pi\,,
 \ee
  which corresponds
 to the  relativistic, field dependent contributions of eq. \eqref{sym-ag1} with $\kappa_0=0$. We  will focus on this system in what follows.
 
  \end{itemize}

  \subsection{The  scalar theory in the extreme relativistic limit }

Action  \eqref{actk01a1} geometrically describes a brane configuration  embedded 
 in a five dimensional space time where the speed of light along the extra dimension, $v_{light}^y\,=\,\kappa_0$,  vanishes since $\kappa_0\to0$: causality would seem to
 require us to select $X>0$ in order to have a well defined square root. Quantities $X>0$ and $X<0$ are respectively space like and time like
 with respect to the four vector $\partial \pi$ relative to the four flat dimensions at $y=const$.  
 
 On the other hand, at the formal level,  the system  allows us to also consider the case where $X<0$, that is a time like scalar derivative $\partial \pi$.
 If $X<0$,  we can  define $\cal N$ and $c$ to be purely imaginary (say,
 ${\cal N}\,=\,i\,\tilde{\cal N}$, $c\,=\,-i\,\tilde c$ with ${\tilde{ \cal N}}$, $\tilde c$ real constants) so to compensate for the imaginary
  `$i$ factor'
   associated with the square root, and get a real action. The action for a time like scalar,  $X<0$, has  the same structure as before: 
   \bea\label{actk01a1a}
  S\,=\,\tilde{\cal N} \int d^4 x\,  \sqrt{-X}  \,
\det{\left[
\delta^\mu_\nu+\, \frac{\tilde c}{ \sqrt{-X}}\left(
\,\delta^\mu_\rho-\frac{\,\partial^\mu \pi  \partial_\rho \pi}{X}
\right)
 \,
 \partial^\rho \partial_\nu \pi
\right]}\, ,
\eea
 and enjoys the same symmetry as eq. \eqref{actk01a1}.  
  A possible geometrical interpretation for this set-up can be found  considering a probe brane embedded in a
   five dimensional space time with  {\it two}
  time  directions. One is the usual $T$, the other is a time like (Wick rotated) version of the extra dimensional coordinate $y$, that we dub $\tilde y$. The five
  dimensional metric to consider  is
  \be
 d s^2\,=\,-\kappa_0^2\,d T^2+\kappa_0^2\,d \vec X^2-d \tilde y^2 
  \ee
 We can define -- analogously as explained in Section \ref{sym-poi} -- an embedding $X^\mu\,=\,x^\mu$, $\tilde y\,=\,\pi$. Calculations can be  carried
on  straightforwardly following the very same steps as  Section \ref{sym-poi}, finding that the action \eqref{sdbi2aaa} leads to action \eqref{actk01a1a}, once
substituting the new  expressions for induced metric and extrinsic curvature. While such geometrical derivation of  action  \eqref{actk01a1a}   can be  useful for determining symmetries and dualities for our system,  
 its
 physical relevance deserves further study,
 since   the physical meaning of {\color{black} bulk} space-times with multiple time directions is  not clear to us. On the
 other hand, let us point out that theories equipped  
 with two time directions in extra dimensions have been considered in string/M-theory contexts, see e.g. \cite{Bars:2000qm}
 and references therein.

 \bigskip
 
 Expanding the determinants in eqs \eqref{actk01a1} and \eqref{actk01a1a} we find four scalar
  Lagrangians in flat space
 \bea
{\cal L}_1&=& \Lambda^2\,\sqrt{|X|} \label{scl1}
\\
 {\cal L}_2&=& \Lambda\left( [\Pi]-\frac{1}{X} [\Phi]\right)
\label{scl2} \\
 {\cal L}_3&=&\frac{1}{\sqrt{|X|}}\left( [\Pi]^2-[\Pi^2] +\frac{2}{X} \left([\Phi^2]- [\Phi] [\Pi] \right)\right)
 \label{scl3} \\
 {\cal L}_4&=&\frac{1}{\Lambda\,X} \left(
[\Pi]^3+2 [\Pi^3]-3 [\Pi^2]  [\Pi] + \frac{3}{X}  \left(2 [\Pi] [\Phi^2]-2 [\Phi^3] - [\Phi] [\Pi]^2+
 [\Phi] [\Pi^2] \right)
\right)\,,
\label{scl4}
\eea
which describe both the cases of $X$ positive or negative.  As before, we use the notation
$\left[ \Pi^n \right] \,=\,{\text{ tr}}\left( \Pi^{n}
 \right) $ and 
 $\left[ \Phi^n\right]\,=\,{\text{ tr}}\left(\partial  \pi  \,\Pi^{n}
\, \partial \pi \right)$. 
We include
an  energy scale $\Lambda$ to make explicit the dimension of each operator.
Each of these four Lagrangians enjoys the scalar symmetry 
\be \label{nsym1}
\delta \pi\,=\,\pi \, w^\mu\partial_\mu \pi\,=\,\frac12 w^\mu\partial_\mu \pi^2
\ee
 with $w^\mu$ an arbitrary constant four vector,
which leaves the action invariant up to boundary terms.
 This transformation lacks the linear `coordinate dependent' part which characterises 
Galileon symmetries (the `$\delta \pi\,=\,w_\mu x^\mu$') hence the system is qualitatively different from Galileons, and we do not reduce to Galileon actions
in any `small derivative'  limit. Additionally, the four actions are also connected by a duality, as discussed in Section \ref{sec-poidbi} (whose
results remain valid in the $\kappa_0\to0$ limit).

\smallskip

Taken by themselves, 
these scalar actions are quite peculiar: there is no limit in which the scalar has standard kinetic terms, since
 standard kinetic terms are not compatible with symmetry
\eqref{nsym1}. Some of these Lagrangians are non analytic, since they contain the square root of $X$, and all of them contain powers of $1/X$. 
   On the other hand, such scalar theories might  make sense when expanded around some background  which solves the equations of motion, or
   by coupling to other fields like gravity.  
  We now discuss a simple, concrete example to develop these points further, and to assess the physical relevance of these systems.

\subsubsection{An explicit example: Part I} \label{exp1}

 In the time-like case $X<0$ these theories seem to have problematic causal properties, which can be fixed by slightly breaking the scalar symmetry (as
 we are going to  discuss now), or by enlarging the system by coupling it to other fields, as 
  dynamical 
 gravity (as we discuss in the next Section, see in particular
 Section \ref{exp2}).   
 
 We  analyse  a concrete  example that is simple, but   illustrative. 
  We consider a linear combination   ${\cal L}_C$ of the Lagrangians \eqref{scl1}-\eqref{scl4} with constant dimensionless coefficients
\be
-{\cal L}_C\,=\,\alpha_1 {\cal L}_1+\alpha_2 {\cal L}_2+\alpha_3 {\cal L}_3+\alpha_4 {\cal L}_4\,.
\ee
We start determining some homogeneous background around which we  expand our theory. Any scalar
configuration which is linear in the coordinates solves the equations of motion.
  This can be checked by direct computation, or by using symmetry arguments. Denote with $\bar \pi\,=\,c_\mu x^\mu$ one scalar configuration, with $c_\mu$ arbitrary vector. Applying
  the scalar transformation  \eqref{nsym1}, we obtain $ \bar \pi+ \delta \bar \pi\,=\,\left( 1+\omega^\rho c_\rho\right)\,c_\mu x^\mu$: so a symmetry
  transformation sends this configuration to an arbitrary other one with a linear
  profile, but with a different vector $c_\mu$. Then, since $\bar \pi=0$ is  a solution, also  
any  $\bar \pi\,=\,c_\mu x^\mu$ must be solution. 
 In order to preserve three dimensional spatial isotropy, 
 we select a background configuration that is linear in time:
   \be \label{linp0}
  \bar{\pi}(t)\,=\,P_0^2\,t
  \ee 
  with $P_0$ an arbitrary constant with dimension of a mass. (We put a $P_0^2$ in the previous formula to assign the correct dimension to the scalar field.)  
We examine 
  the
  dynamics of fluctuations around
    $\bar{\pi}(t)$:
\be\pi(t,\vec x)\,=\,\bar{\pi}(t)+\hat \pi(t,\vec x)\,. \label{ans0}
\ee
 An homogeneous           background  $\bar{\pi}(t)$
spontaneously breaks the symmetry \eqref{nsym1} down to a residual symmetry
\be\label{sb-sym}
\delta \pi\,=\,\pi \, w^i\partial_i \pi\,,
\ee
with $w^i$ an arbitrary {\it three} spatial vector. Indeed, transformation \eqref{sb-sym} leaves invariant  any   function $\bar{\pi}(t)$,
and only acts on the fluctuations $\hat \pi(t,\vec x)$ introduced
in eq. \eqref{ans0}. In the limit of small fluctuations, the residual symmetry \eqref{sb-sym} acts at linear order on $\hat \pi(t,\vec x)$
as
\be \label{sb-sym2}
 \delta \hat \pi(t,\,\vec x)\,=\,\bar \pi(t) \, w^i\partial_i \hat \pi(t,\,\vec x)\,.
 \ee 

Expanding the  combination ${\cal L}_C$  at quadratic order in small fluctuation $\hat \pi$ around the background $\bar{\pi}(t)$, we do not find a standard kinetic term
 for the scalar fluctuation, but instead the quantity
\be \label{quad-fl-lag} 
{\cal L}^{quad}\,=\, q(t) \,( \vec{\nabla} \hat \pi )^2
\ee
 where $q(t)$ depends on $\bar{\pi}(t)$, and on the coefficients $\alpha_i$ characterising the combination ${\cal L}_C$ we selected. Such
 quadratic Lagrangian for fluctuations only contain spatial derivatives of $\hat \pi$, and  lacks the time derivative piece $\dot {\hat \pi}^2$. This fact is  easier to understand 
 in terms of the residual symmetry \eqref{sb-sym2}:
 while the quadratic Lagrangian  \eqref{quad-fl-lag} is invariant under this transformation (up to boundary terms), a term like $\dot {\hat \pi}^2$ is not. 
  The  system described by the quadratic Lagrangian \eqref{quad-fl-lag} is degenerated, and does not satisfy the conditions of Leray's theorem for having a well defined Cauchy's problem \cite{Bruneton:2006gf} (healthier
   degenerate systems will be discussed in the next section, when coupling with gravity). 
  
  A way  out is to break {\it explicitly} symmetry \eqref{nsym1}, for example by including  a standard kinetic term with small overall
  coefficient. We add such a  term to our Lagrangian
  \be\label{lag-wkt}
-  {\cal \tilde L}_C\,=\,\alpha_0 X+ \alpha_1 {\cal L}_1+\alpha_2 {\cal L}_2+\alpha_3 {\cal L}_3+\alpha_4 {\cal L}_4\,. 
  \ee
  The first term proportional to $\alpha_0$ breaks symmetry  \eqref{nsym1}; in this case, again our
    homogeneous solution $\bar{\pi}(t)$ 
    of eq. \eqref{linp0} solves background equations of motion, since the term proportional to $\alpha_0$ has a Galilean symmetry. 
     Studying the dynamics of quadratic fluctuations, associated with this Lagrangian, we find a healthy kinetic term
  for the fluctuation $\hat \pi$ at quadratic level, if $\alpha_{0,1}$ are positive: 
  \bea \label{lag111}
  {\cal \tilde L}^{quad}&=& \alpha_0\,\left(\dot{\hat \pi}^2-c_\pi^2\, \partial_j \hat \pi \,\partial^j \hat \pi \right)\,,
\\
c_{\pi}^2&=& 1-\frac{\alpha_1\,\Lambda^2}{2\,P_0^2\,\alpha_0} \label{defcpi}
 \,. \eea
  By an appropriate choice of the quantities, $\alpha_{0,\,1}$ and $P_0$, we can ensure that
   \be
   0\,<\,c_\pi\,\le\,1
\label{int-cpi}
\,,
   \ee
    so that fluctuations get healthy kinetic terms. We can also 
   then canonically normalize the field 
   \be \label{canof1}
   \hat \pi \,\to\,\frac{1}{\sqrt{2\,\alpha_0}}\hat \pi  
   \,,
   \ee
   so to have a canonical kinetic terms, with a speed of sound different from unity. In order to have a consistent system when
   $\alpha_0$ is small -- with $0< c_\pi\le 1$ -- we have to require 
   that the energy scale of the background solution $P_0$ is larger than the scale
   entering 
    $\Lambda$ in the Lagrangian.

  We can also proceed and examine higher order self interactions for the fluctuations. We limit to turn
  on the coefficients $\alpha_0$, $\alpha_1$ in eq. \eqref{lag-wkt}, while setting to zero the remaining $\alpha_i$.
      After canonically normalize the field, as in eq. \eqref{canof1}, we find
  that the Lagrangian expanded up to fourth order results
  \be \label{lag-up-4}
\tilde  {\cal L}
 \,=\, \frac12 \left[ \dot{\hat \pi}^2-c_\pi^2\,( 
\nabla \hat \pi )^2  \right]-\frac{\sqrt{\alpha_0} \,\left(1-c_\pi^2 \right)^2}{\sqrt{2}\,\alpha_1\,\Lambda^2}\,\dot{\hat \pi}\,( 
\nabla \hat \pi )^2\,+ \frac{\left(1-c_\pi^2\right)^3\, \alpha_0\,( 
\nabla \hat \pi )^2\, \left(4 \dot{\hat \pi}^2+  ( 
\nabla \hat \pi )^2
 \right)}{4\,\alpha_1^2\,\Lambda^4}+\dots
  \ee
  with $c_\pi$ given in eq. \eqref{defcpi}. As long as the sound speed lies in the interval \eqref{int-cpi}, the system is defined also in a
  small $\alpha_0$ regime, since interactions are suppressed.  
  
  \smallskip
  On the other hand, symmetry \eqref{sb-sym} is explicitly broken by the contribution of the kinetic term. 
  If we would like to recover the symmetry, in a regime where $\alpha_0$ is very small, we need to go outside the safe interval   \eqref{int-cpi}
  for the sound speed. Consider Lagrangian \eqref{lag-up-4} in a regime 
    where $\alpha_0\to 0$, $c_\pi \to \infty$, such that the
 combination $\alpha_0\,c_\pi^2\,=\,const\equiv \,2\,\beta\,\alpha_1^2$ for some arbitrary constant $\beta$. Moreover, let us rescale $\hat \pi\,\to\,\hat \pi/c_\pi$.   In such limit, the Lagrangian \eqref{lag-up-4} becomes
 \be
 \tilde  {\cal L}_{sym}
 \,=\, 
 -\frac12\,( 
\nabla \hat \pi )^2
-\frac{\sqrt{\beta} }{\Lambda^2}\,\dot{\hat \pi}\,( 
\nabla \hat \pi )^2\,- 
\frac{\beta}{2\,\Lambda^4} 
\,{( 
\nabla \hat \pi )^2\, \left(4 \dot{\hat \pi}^2+  ( 
\nabla \hat \pi )^2
 \right)}+\dots
 \ee
 and each term respect symmetry  \eqref{sb-sym} at 
 the corresponding order in 
 perturbations. 
 
   \smallskip
   
It would be interesting to analyse whether the symmetries and the properties  of our scalar actions can protect their structure under  corrections, for example 
 against scalar self loops, leading to non-renormalization theorems as
it happens for Galileons.  In fact, this kind of questions have been recently reconsidered 
for a wide class of derivatively coupled theories \cite{Goon:2016ihr}, using simple yet
powerful methods based on power counting techniques (see e.g. \cite{Burgess:2003jk} for a review). 
The scalar transformation that we consider -- eq. \eqref{nsym1}  -- is a  symmetry of the action only up to boundary terms. Usually set-ups with this property are protected under quantum corrections, as discussed in \cite{Goon:2016ihr}. Breaking
  it spontaneously by  selecting a non-trivial homogeneous scalar background should not spoil these features. An explicit symmetry breaking (as done
  by adding a kinetic term to the Lagrangian in eq. \eqref{lag-wkt}) might still lead to a system where corrections can be kept under control: in the limit
  in which the explicit symmetry breaking parameters are small (for our previous example, $\alpha_0\ll\alpha_1$), one expects  quantum corrections to be
  small, at most proportional to $\alpha_0$ and its powers
   \cite{'tHooft:1979bh}. It would be interesting to concretely develop these arguments in our specific example, where we know that, in the limit
   of $\alpha_0$ small, the background profile
   $P_0/\Lambda$ must be  large.  We will return to these issues from a different perspective in the next Section \ref{sec-est}, where we
  will study the related topic of what happens to our scalar system when coupling with gravity.

\subsection{A generalization} \label{sec-gen}
If we consider a field redefinition
  \be
  \pi \to f(\pi)
  \ee
  and apply it to action \eqref{actk01a1a}, 
 we obtain a new action  which explicit depends on $\pi$ (and not only on its derivatives) 
 thanks to an overall factor in front of the determinant.  It  is given by
  
 \bea\label{actk01a3}
 S\,=\,{\cal N} \int d^4 x\, f'(\pi)  \sqrt{X}  \,
\det{\left[
\delta^\mu_\nu+\,c \frac{1}{ \sqrt{X}}\left(
\,\delta^\mu_\rho-\frac{\,\partial^\mu \pi  \partial_\rho \pi}{X}
\right)
 \,
 \partial^\rho \partial_\nu \pi
\right]}\, .
\eea
  Such action satisfies a  symmetry which generalizes \eqref{nsym1}, and
is given by
\be
\delta \pi\,=\,f(\pi)\,w^\mu\partial_\mu \pi
\ee
for constant vector $w^\mu$. As a byproduct, 
 this fact  implies that the equations of
motion associated with action \eqref{actk01a3} are invariant under constant rescaling of the field
  $\pi$: $ \pi \to \lambda \pi $. This since in this case $f\,=\,\lambda \pi$, $f'\,=\,\lambda$, and the constant $\lambda$ goes in front of 
  the integral in eq. \eqref{actk01a3}, without affecting the equations of motion. 

\subsection{Extreme relativistic limit of DBI Galileons in AdS space}
 
  All what we said so far can be straightforwardly extended to the case of DBI Galileons embedded in AdS
  space, which is another system with an interesting geometrical interpretation (see Section \ref{sec-adsdbi}). Taking the $\kappa_0\to 0$ limit of action \eqref{acaddb1}, we get
   \bea\label{actk01d}
 S\,=\,{\cal N} \int d^4 x\, e^{-3 \pi/\ell} \,  \sqrt{X}  \,
\det{\left[
\delta^\mu_\nu+\,c \frac{e^{ -\pi/\ell}}{ \sqrt{X}}\left(
\,\delta^\mu_\rho-\frac{\,\partial^\mu \pi  \partial_\rho \pi}{X}
\right)
 \,
 \partial^\rho \partial_\nu \pi
\right]}\, .
\eea
  This action  is symmetric under the field-dependent transformation (the $\kappa_0\to0$ limit of eq. \eqref{scsads1})
\be \label{symk01a}
\delta \pi\,=\, \frac{\ell}{2}\left( e^{2 \pi/\ell}-1\right)\,w^\mu \partial_\mu \pi\, ,
\ee
for arbitrary constant vector $\omega^\mu$.  Again,  this transformation lacks the linear `coordinate dependent' part which characterises 
Galileon symmetries (the `$\delta \pi\,=\,w_\mu x^\mu$'); hence the system is qualitatively different from Galileons.   
  Also for the AdS case, the set-up admits a simple
      generalisation: the structure of action \eqref{actk01d} is the
      same by doing a field redefinition 
\begin{equation}
 \pi \to -\ell \,\ln{h}(\pi), 
\end{equation} 
for arbitrary
      function $h$. The action becomes
\begin{equation}
\label{actk01aa}
 S\,=\, \int d^4 x\,h^3\,{h'}\,\sqrt{X}  \,
\det{\left[
\delta^\mu_\nu+\,c \frac{h}{ \sqrt{X}}\left(
\,\delta^\mu_\rho-\frac{\,\partial^\mu \pi  \partial_\rho \pi}{X}
\right)
 \,
 \partial^\rho \partial_\nu \pi
\right]}\, .
\end{equation}
The associated symmetry becomes
\be \label{symk01a}
\delta \pi\,=\, \frac{\ell}{2}\left( \frac{1}{h^2}-1\right)\,w^\mu \partial_\mu \pi\,.
\ee 
The resulting action and symmetry, eqs \eqref{actk01aa} and \eqref{symk01a},
 are similar, although not identical, to the system discussed in the previous subsection \ref{sec-gen}.

\smallskip
  
\section{Minimal coupling with gravity: degenerate scalar-tensor theories}
\label{sec-est}

Our flat space `extreme relativistic' Lagrangians with $\kappa_0\to0$ can be minimally coupled to gravity in a consistent way, by promoting   
the flat four dimensional slices to arbitrarily curved slices with dynamical four dimensional metric.  This
 relates our systems to beyond Horndeski \cite{Gleyzes:2014dya,Gleyzes:2014qga} and extended scalar-tensor theories \cite{Langlois:2015cwa,Crisostomi:2016czh,Achour:2016rkg}, providing a geometrical perspective to the latter systems. 

\subsection{From DBI Galileons to beyond Horndeski theories}

We can minimally couple with gravity the   extreme relativistic  actions \eqref{actk01a1} and \eqref{actk01d}, promoting the flat
four dimensional metric tensor $\eta_{\mu\nu}$ to a dynamical tensor $q_{\mu\nu}$, and writing respectively
\begin{equation}
\label{actk01bA}
 S\,=\,{\cal N}\, \int d^4 x \, \sqrt{-q}\, \sqrt{X}  \,
\det{\left[
\delta^\mu_\nu+\,c \frac{1}{ \sqrt{X}}\left(
\,\delta^\mu_\rho-\frac{\,\partial^\mu \pi  \partial_\rho \pi}{X}
\right)
 \,
 \nabla^\rho \partial_\nu \pi
\right]}\,.
\end{equation}
and 
\begin{equation}
\label{actk01b}
 S\,=\,{\cal N}\, \int d^4 x\, e^{-3 \pi/\ell} \, \sqrt{-q}\, \sqrt{X}  \,
\det{\left[
\delta^\mu_\nu+\,c \frac{e^{-\pi/\ell}}{ \sqrt{X}}\left(
\,\delta^\mu_\rho-\frac{\,\partial^\mu \pi  \partial_\rho \pi}{X}
\right)
 \,
 \nabla^\rho \partial_\nu \pi
\right]}\,.
\end{equation}
Eq. \eqref{actk01b}   reduces to \eqref{actk01bA} in the limit of infinite AdS radius $\ell\to\infty$. For simplicity, in what follows 
we focus on the Poincar\'e limit $\ell\to\infty$, although similar considerations can be done for the AdS case as well.

Interestingly, the scalar-tensor theories one obtains by expanding the determinants in the previous expressions are consistent, although the associated EOMs are generally of higher order. The set of Lagrangians one finds corresponds  to Lagrangian densities \eqref{scl1}-\eqref{scl4}, with standard 
derivatives replaced by covariant derivatives. 
 Such scalar-tensor theories  belong to the
class of beyond Horndeski theories \cite{Gleyzes:2014dya,Gleyzes:2014qga}. 
It is easier to  check  this fact 
 using the idempotent `projection tensor' 
 \be \label{proten}
 P^\mu_\nu\,=\,\delta^\mu_\nu-\frac{\,\nabla^\mu \pi  \nabla_\nu \pi}{X}\, ,
 \ee
 which satisfies the relation $ P_\mu^\nu\,\nabla_\nu \pi\,=\,0$.
Using this quantity, it is possible to prove (see  section IIB of \cite{Crisostomi:2016tcp}) that the theories of beyond Horndeski 
can also be expressed in terms of a determinantal expression. They 
can be 
 written as
 \be \label{byhact}
 S\,=\,\int d^4 x \sqrt{-q}\,A(\pi,X) \,\det{\left[ \delta^\mu_\nu+B(\pi,X) P_{\nu}^\rho\,\nabla_\rho \partial^\mu\,\pi\right]}\,,
 \ee
 where $A$, $B$ are arbitrary function of $\pi, X$, and $\nabla$ denotes covariant derivative with respect to a 4d metric $q_{\mu\nu}$. Although these theories are characterised by EOMs of  order higher than two, they propagate at most
 three dofs. 
 Action \eqref{actk01b} belongs to this class of theories: hence it does not propagate more than three
 degrees of freedom. We emphasize that actions as \eqref{byhact} do not need supplementary gravitational counterterms for being consistent. This fact
 relates a limit of DBI Galileons with beyond Horndenski.  
  
  \bigskip
  
  We can also investigate geometrically the covariantization procedure for the extreme relativistic limit of DBI Galileons, 
  in terms of a probe brane in an five dimensional set-up. 
Recall  that for studying 
  Poincar\'e  DBI Galileons we consider a
 five dimensional metric  as \be \label{metrd5a} ds^2_{(5)}\,=\,
\kappa_0^2\,\eta_{\mu\nu}\,d X^\mu d X^\nu+d y^2 \ee i.e. the four dimensional           
slices $y\,=\,$constant have flat  metric.  We discuss here the possibility of
promoting the metric of 4d slices to a dynamical field, writing \be \label{metrd5b}
ds^2_{(5)}\,=\kappa_0^2\,q_{\mu\nu}\,d X^\mu d X^\nu+d y^2 \ee with
$q_{\mu\nu}$ a dynamical tensor. We choose again the usual foliation associated with a
brane embedded on this geometry, $
X^\mu \,=\, \kappa_0\, x^\mu$,
$
y \,=\, \pi  
$.
 The induced metric on the brane is 
\begin{equation}
g_{\mu\nu}\,=\,\kappa_0^2\,q_{\mu\nu}+\partial_\mu \pi \partial_\nu \pi \, ,
\end{equation}
with determinant
\begin{equation}
\sqrt{-g}\,=\, \kappa_0^3\, \sqrt{-q}\,\sqrt{\kappa_0^2+X}\,,
\end{equation}
where 
\be X=q^{\mu\nu}\nabla_\mu\pi\nabla_\nu\pi\,.\ee
 Its inverse is (4d indexes are raised with $q^{\mu\nu}$)
\be
g^{\mu\nu}\,=\,\frac{1}{\,\kappa_0^2}\,\left(q^{\mu\nu}-\frac{\partial^\mu \pi\, \partial^\nu \pi}{\kappa_0^2\,+X} \right)\,.
\ee
The extrinsic curvature tensor is
\be
 K_{\mu\nu}\,=\,-\frac{\kappa_0}{\sqrt{\kappa_0^2 +X}}\,\left( \nabla_\mu \nabla_\nu \pi
\right)\, .
\ee

We now consider the extreme relativistic limit
\be
\kappa_0\to0
\,.
\ee

Most of the geometrical quantities written above become singular in this limit: or they vanish, or they become infinite. On the other hand,
  the determinantal action \eqref{acaddb1} 
  \bea \label{lagadsdbiAA}
{\cal S}\,=\, \frac{\cal N}{\kappa_0^3}\int d^4 x\, \sqrt{-g} 
\det{\left[
\delta_\mu^\nu-\,c\,\kappa_0\, g^{\mu \alpha}\,K_{\alpha \nu}
\right]}\, ,
\eea  
  as a whole has a smooth limit,  since taking $\kappa_0\to0$ we obtain the regular expression
\begin{equation}
\label{actk01bAa}
 S\,=\,{\cal N}\, \int d^4 x \, \sqrt{-q}\, \sqrt{X}  \,
\det{\left[
\delta^\mu_\nu+\,c \frac{1}{ \sqrt{X}}\left(
\,\delta^\mu_\rho-\frac{\,\partial^\mu \pi  \partial_\rho \pi}{X}
\right)
 \,
 \nabla^\rho \partial_\nu \pi
\right]}\,,
\end{equation}
which indeed coincides with action \eqref{actk01bA}. 
So we learn that when taking the  limit $\kappa_0\to0$, a  brane 
action built with appropriate combinations of geometrical quantities leads to  sensible scalar-tensor theories.
  We then find a connection between certain DBI Galileons and a special case of beyond  Horndeski theories.

   \subsection{The duality}

   We now discuss a way to extend the duality transformation presented in Section \ref{dua-poi} to the case
   of curved space time. Our aim is to 
    proceed 
    as much as possible along the same steps we followed in discussing duality with    non-dynamical gravity.    When coupling the scalar with dynamical  gravity, however, defining the action of a duality is a very delicate matter, as already pointed
   out for the case of standard Galileons in \cite{deRham:2014lqa} (see also \cite{Baratella:2015yya}). The main issue is how to define the transformation of the dynamical metric
   $q_{\mu \nu}(x)$ under duality. We study here a particular case of duality, which is nevertheless sufficient for finding 
   a novel relation among the different Lagrangians of eqs  \eqref{scl1}-\eqref{scl4}, once they are minimally coupled with gravity. 
   
   Consider the  field dependent  map among two sets of coordinates
\be\label{trdbig}
x^\mu\,\,\Rightarrow\,\,
\tilde x^\mu\,=\,x^\mu+\frac{1}{\sqrt{X}} \,\partial^\mu \pi\,,
\ee
(where, as for the case of Galileons,   we choose units which set to one the parameter  $\Lambda_S$ which is 
 needed for dimensional reasons.)  Recall that $X\,=\,q^{\mu\nu} \,\nabla_\mu \pi\,\nabla_\nu \pi$.
 
 The duality is defined as the transformation which maps the line element $d x^\mu$ of the first set of coordinates, with
 the line element $d \tilde x^\mu$ of the second set of coordinates as follows
 \be\label{trdbig1}
 d x^\mu\hskip0.5cm \Rightarrow \hskip0.5cm
d \,
\tilde x^\mu\,=\,d\,x^\mu+d\,\left(
\frac{1}{\sqrt{X}} \,\partial^\mu \pi \right)\,=\,
d\,x^\mu+\nabla_{\nu}\left( \frac{1}{\sqrt{X}} \,\partial^\mu \pi \right)\,d x^\nu
\,.
\ee
{\color{black} The transformation matrix between these line elements is }
\bea
J^\mu_\nu&\equiv&\frac{d \tilde x^\mu}{d x^\nu}   \,=\, \delta^\mu_\nu+ \nabla_\nu \left( \frac{1}{\sqrt{X}} \partial^\mu \pi\right)\,.
\eea 
We demand that the  dual metric $\tilde q_{\mu\nu}(\tilde x)$   is a {\it scalar} under duality:
\be \label{dume}
\tilde q_{\mu\nu}(\tilde x)\,=\, q_{\mu\nu}( x)\,.
\ee
We also demand that there exists a dual scalar field $\tilde \pi$, which we can use for mapping back the coordinates $\tilde x^\mu$ to $x^\mu$:

\be\label{trdbig2}
\tilde x^\mu\,\,\Rightarrow\,\,
 x^\mu\,=\,\tilde x^\mu-\frac{1}{\sqrt{\tilde X}} \,\tilde \partial^\mu \tilde \pi\,,
\ee
Comparing eqs \eqref{trdbig2} and \eqref{trdbig}, and using \eqref{dume}, we find that the simplest way to satisfy our conditions is to impose that 
the derivative of $\pi$ is a scalar under duality
\be \label{cod1ab}
\tilde \partial^\mu \tilde \pi (\tilde x)\,=\,\,\partial^\mu \pi (x)
\,.
\ee
This is  analogous to what we have seen in the case of flat metric;  also, these results imply that the {\it induced} four dimensional metric 
$g_{\mu\nu}$ is scalar under duality (indexes are raised/lowered with curved metric $q_{\mu\nu}$)
\be
g_{\mu\nu}(x)\,=\,\kappa_0^2 q_{\mu\nu}+\partial_\mu \pi \partial_\nu \pi \,\,\,\Rightarrow\,\,\,
\tilde g_{\mu\nu} (\tilde x) \,=\, g_{\mu\nu}(x) \,,
\ee
as happens in flat space.

Eq. \eqref{cod1ab} implies 
 
 \be
 \tilde \partial_\mu \tilde \pi\, d \tilde x^\mu =\,\,\partial_\mu \pi\,J^\mu_\nu\,d x^\nu\,=\,\partial_\mu \pi\, d x^\mu+\partial_\mu \pi\, \nabla_\nu
 \left( \gamma \partial^\mu \pi\right)d x^\nu\,.
 \ee

 Integrating both sides of the previous relation, we find a non local  relation among the field $\pi$ and its dual
  \be
  \pi(x)\,\Rightarrow\, \tilde \pi(\tilde x)\,=\, \pi(x)+ \int^x\,d \bar x^\lambda\,\nabla_\lambda\left(\gamma\partial^\rho \pi\right)\,\partial_\rho \pi
 \,.\ee
The results so far imply that, under the action of the duality, 
 \bea
 d^4 x&\Rightarrow&  d^4 \tilde{x}\,=\,\left(\det{J}\right)\, d^4 x \,=\, \det{\left[\delta^\mu_\nu+ \nabla_\nu \left( \frac{1}{\sqrt{X}} \partial^\mu \pi\right)\right]}\, d^4 x  \,,\\
X&\Rightarrow&   \tilde X \,=\, X\,.
 \eea
We meet a serious  problem however, with respect to  how to define in a consistent way the dual of the second derivative of the scalar field. This since
the expression $\nabla_\mu \,\partial^\nu \,\pi$ contains a covariant derivative, which does not transform properly as a vector, because the metric does not transform as a tensor under duality. Hence, following this route, we can not define a dual version
 of the entire set of scalar-tensor actions we examined. 

Less ambitiously, we 
can nevertheless  define a dual  version of the action
\bea
S&=&\int d^4 x \,\sqrt{-g}
\\
&=&
\int d^4 x \,\sqrt{-q}\,\sqrt{X}
\eea
 associated with Lagrangian \eqref{scl1} minimally coupled with gravity. It is
 \bea
 \tilde S&=&\int d^4 \tilde x \,\sqrt{-\tilde q}\,\sqrt{ \tilde X}
 \\
 &=&\int d^4  x \,\sqrt{- q}\,\sqrt{  X}\, \det{\left[\delta^\mu_\nu+ \nabla_\nu \left( \frac{1}{\sqrt{X}} \partial^\mu \pi\right)\right]}\
\,, \eea
 which is a particular case of eq. \eqref{actk01bA} for $c=1$. Expanding the determinant, we find a minimal coupling
 with gravity of all the actions  \eqref{scl1}-\eqref{scl4}, with fixed coefficients: our duality maps  scalar and metric fields in such a way
 to generate all our actions starting from the simplest among them. 
   It would be interesting  to extend these findings to determine  an action of the duality 
   which applies to all the scalar-tensor actions we have studied.

\subsection{The symmetry}
 
We can now  ask about the fate of the flat space symmetries we discussed  in the previous Sections. 
In the presence of dynamical  gravity, as in  action \eqref{actk01bA}, we normally break the scalar
symmetry 
\be
\delta \pi\,=\,\pi \,w^\mu \partial_\mu \pi \label{symk01f}
\ee
since  the equations of motion for the dynamical metric field are not necessarily invariant under such transformation.
   On the other hand,
two classes of  general  arguments can be made.  The first set of considerations concerns systems
in which gravity is still not dynamical, but with a non trivial fixed metric $\bar q_{\mu\nu}$ on the 4d slices.
 If such space  5d space time 
 has still some isometries, it is possible to use
the techniques of \cite{Goon:2011uw,Goon:2011qf} for constructing a scalar transformation
which generalises \eqref{symk01f} and is a symmetry of the action.    {If, on the other hand,  the  5d
  space time     does not admit any isometry, it might still be possible to describe it as a
  `small perturbation'  of some symmetric   space time. 
 A second kind of considerations can be made when gravity becomes dynamical, and the metric
 $q_{\mu\nu}(x)$ on the 4d slices is a dynamical field with its own equations of motion. In this set-up, the
 equations of motion for $q_{\mu\nu}$ normally break the scalar symmetry \eqref{symk01f}. In some situations,
the symmetry 
  could be  broken by
  gravity  in a soft way, and 
  some of its properties might  be maintained}. Such
arguments have been used in phenomenological approaches of Galileons to cosmology,
see e.g. \cite{Burrage:2010cu,Pirtskhalava:2015nla,Gratia:2016tgq}. 
 Moreover, 
as
   advocated for example in
\cite{deRham:2010eu},   one could try to  promote the global 5d Poincar\'e and AdS symmetries
considered so far to local symmetries, and analyse their physical 
consequences for the   brane induced action. 
  It
would be    interesting   
    to develop these  considerations by studying concrete systems, to explicitly understand the fate of
     symmetries when gravity is turned on. 
       In the present context, we limit ourselves  to reconsider the 
  simple, explicit example of Section \ref{exp1}, for understanding the  behavior
    of fluctuations when the set-up is coupled with 
    dynamical 
    gravity, and at what extent the symmetry is broken.
  
  
\subsubsection{An explicit example: Part II}
\label{exp2}

We focus on time like systems with $X<0$, and reconsider the example of Section \ref{exp1} in the present context. 
We analyse the dynamics of fluctuations around a time dependent homogeneous background $\bar \pi(t)$ which solves the equations
of motion for scalar and metric fields. We have seen in Section \ref{exp1} that, in absence of gravity, a residual symmetry prevents us from having canonical kinetic
terms for scalar fluctuations around
 our background profile.
 When gravity is turned on, the situation can be improved. The symmetry gets dynamically  broken by
gravitational effects, and fluctuations acquire kinetic terms,  thanks to a
 kinetic 
 mixing among   the scalar and  the metric sectors. Let us see  these facts more explicitly. 

We focus  
for simplicity 
on a scalar-tensor action based of Lagrangian density ${\cal L}_3$ of eq. \eqref{scl3}, to which we add an Einstein-Hilbert (EH) term
\bea
 {\cal S}_3&=&\int\,d^4x\,\sqrt{-q}\,\left[\mu^2\,R-
 \frac{\alpha_3}{\sqrt{|X|}}\left( [\Pi]^2-[\Pi^2] +\frac{2}{X} \left([\Phi^2]- [\Phi] [\Pi] \right)\right)\right]
 \label{scl3a}\,,
\eea
 where all  derivatives are covariant derivatives,  $\mu$ is a constant with dimension mass, and $\alpha_3$
 a dimensionless constant.  An EH term is included since it does not break the symmetry further than what is done by the covariant derivatives in \eqref{scl3a}: it will
 be nevertheless
  important when discussing fluctuations. An EH term does not introduce ghosts in this case: this since such term belongs to quartic Horndeski  theories, it can be merged with no harm with the
  combination proportional to $\alpha_3$, which belong to quartic beyond Horndeski  (see e.g. \cite{Gleyzes:2014qga,Langlois:2015cwa,Crisostomi:2016tcp} 
  for details and related discussions).
    The resulting theory is disformally related to Horndeski, but
  only in absence of matter; we  do not discuss here such disformal transformation.  
  
 We are interested to study  configurations which admit Minkowski space as metric background.
 Einstein equations, when evaluated on a Minkowski background,
 impose
  the following condition on the scalar field: 
\begin{eqnarray}
0& = &
\frac{1}{2} \eta_{\mu\nu}\left( [\Pi]^2 - [\Pi^2] + 2 \pi^{,\alpha}[\Pi]_{,\alpha}  \right) -  \pi_{,\alpha\mu\nu}\pi^{,\alpha}-[\Pi]\Pi_{\mu\nu}  \nonumber \\
& &  + \frac{1}{X}\left[\phantom{\frac{}{}} -\eta_{\mu\nu}\left([\Pi][\Phi]+\pi^{,\alpha}\pi^{,\beta}\pi^{,\gamma}\pi_{,\alpha\beta\gamma} + 3 [\Phi^2]   \right) + [\Phi]\Pi_{\mu\nu}  + 2 \pi^{,\alpha}\pi^{,\beta}\Pi_{\alpha\mu}\Pi_{\beta\nu}  \right. \nonumber \\ 
& &  \left.+ 2\pi^{,\alpha}\pi^{,\beta}\pi_{,\alpha\beta(\nu}\pi_{\mu)} + 2 \pi^{,\alpha}\Pi_{\alpha\beta}\Pi^\beta{}_{(\mu}\pi_{,\nu)}-\pi^{,\alpha}\pi_{,\mu}\pi_{,\nu}[\Pi]_{,\alpha} +\frac{1}{2} ([\Pi]^2 - 3 [\Pi^2] )\pi_{,\mu}\pi_{,\nu}  \right]\nonumber \\
& & + \frac{1}{X^2}\left[ 3 g_{\mu\nu} [\Phi]^2 - 6 [\Phi]\pi^{,\alpha}\Pi_{\alpha(\mu}\pi_{,\nu)} 
+ 3 [\Phi^2]\pi_{,\mu}\pi_{,\nu} \right] \,. \label{condgr1}
\end{eqnarray}
 Such condition breaks the scalar symmetry as in \eqref{symk01f}. On the other hand
 such system of equations still admits a   scalar solution
which is linear in the coordinates in Minkowski space:
\be
\pi\,=\,c_\mu x^\mu
\ee
for arbitrary constant vector $c_\mu$. This because such configuration satisfies the scalar equation (as explained
  Section \ref{exp1}) and
at the same time satisfies condition \eqref{condgr1} (because each of its terms contain second derivatives, hence it vanishes when evaluated on
 a linear scalar  configuration).  It 
 would be interesting to investigate whether
  this fact can be  associated with some remnant of a symmetry.
 Hence we are allowed to select a  
 time dependent homogeneous profile 
 \be\label{linp02}
 \pi_0(t)\,=\,P_0^2 t\ee  (see eq. \eqref{linp0}) as background  scalar solution,
  with $P_0$ is some arbitrary parameter  with dimension of a mass. 

   We study the dynamics of fluctuations (scalar, tensor) around this scalar profile and Minkowski
  space.
   Scalar and tensor fluctuations are defined at linear  level around our background as: 
 \bea
 q_{\mu\nu}\,d x^\mu d x^\nu&=&-\left(1+2 N\right)\, d t^2+2\,B_{,i}\,d x^i \,d t+\left[\left(1+2 \zeta\right)\delta_{ij}+h_{ij} \right]\,d x^i\,d x^j
 \,,\\
 \pi&=&P_0^2\,t+\hat \pi
\,.
 \eea
Here $N$, $B$ are the standard ADM constraints, $\hat \pi$, $\zeta$ scalar fluctuations, and $h_{ij}$ transverse traceless tensor fluctuations. 
Constraint equations impose the following relations
\bea
N&=&
0\,,
\\
B&=&\frac{2 \,\alpha_3\,\hat \pi}{\mu^2+2 P_0^2 \alpha_3}+\psi \hskip1cm {\text{with}}\hskip1cm \nabla^2\psi\,=\,
\frac{6\,\alpha_3\,\ddot{\pi}}{
\left( \mu^2+2 P_0^2\,\alpha_3\right)}
\,,
\\
\dot{\hat \pi}&=&-\frac{\mu^2+2 P_0^2 \alpha_3}{2 \alpha_3\,}\zeta
%
\,.
\eea
After
imposing the constraints, we find that the quadratic Lagrangian
for scalar fluctuations contain only one propagating mode, $ \zeta$, whose 
quadratic Lagrangian is
\be
{\cal L}^{(2)}_{\zeta}\,=\,6\alpha_3\,P_0^2\left[
\dot {\zeta}^2
-\,\left(\frac{\mu^2+2 \alpha_3 P_0^2}{3\,\alpha_3 P_0^2} \right)
\left( \partial_i  \zeta \right)^2
\right]\,.
\ee
The mode still propagates in the limit $\mu\to0$: the kinetic mixing among scalar fluctuations and the constraints, induced by the covariant
derivatives in action\eqref{scl3a} is sufficient for fully breaking the symmetry and give dynamics to scalar fluctuations. In order to avoid
ghosts, one imposes $\alpha_3>0$. 

The quadratic action for tensors, on the other hand, results
\be
{\cal L}^{(2)}_{h}\,=\,
\frac{\mu^2+\alpha_3\,P_0^2}{4}
\left(
{\dot {h}_{ij}^2}
-  \frac{\mu^2}{\mu^2+ \alpha_3 P_0^2}
\left( \partial_l{h}_{ij}\right)^2
\right)
\,.
\ee
In order for tensors to propagate with no strong coupling issues (as 
  pointed out in 
 \cite{deRham:2016wji,BenAchour:2016ftp}), we need $\mu\neq0$.  The resulting system propagates
three healthy degrees of freedom. Notice that both the sound speeds $c_\zeta$, $c_h$ are less than one,
if $\alpha_3$, $\mu^2$ are positive quantities.    It would be interesting to understand whether gravity 
breaks the scalar symmetry \eqref{symk01f} in some spontaneous way, and whether (around Minkowski space)
 there are some remnants of the scalar symmetry
that are also a  symmetry of the gravitational equations of motion \eqref{condgr1}. We plan to investigate
this subject in a separate publication.

 \subsection{The relation  with a broader class of  EST theories}

 Beyond Horndeski are not the only scalar-tensor theories, with higher order equations of
 motion, which are made consistent by the existence of primary constraints preventing the
 propagation of additional degrees of freedom. Generalisations of this case have been
 studied  recently, and have been dubbed EST \cite{Crisostomi:2016czh} or DHOST \cite{Achour:2016rkg},
 using an approach developed by Langlois and Noui \cite{Langlois:2015cwa}.
  In particular, in \cite{Crisostomi:2016czh} it has been pointed out
 that a class of consistent extensions of beyond Horndeski theories can be built in terms of the
 projection operators $P_\mu^\nu$ introduced in \eqref{proten}. 
 
We introduce the two index tensor
\begin{equation}
Q_\mu^\nu\,\equiv\, P_\mu^\alpha \nabla_\alpha \nabla^\nu \pi
\,,
\end{equation}
and consider a   scalar-tensor
theory described by an action which is a combination of scalar quantities formed with $Q_{\mu\nu}$, 
like
\begin{equation}
S\,=\,\int  d^4 x\,\sqrt{-q}\,\Big[ A_1(\pi,X)\,Q_\mu^\mu+
A_2(\pi,X)\,\left(Q_\mu^\mu\right)^2+ 
A_3(\pi,X)\,\left(Q_\mu^\nu \,Q_\nu^\mu\right)+\dots\Big] \label{est-a1}
\end{equation}
for arbitrary functions $A_i$.  
Thanks to  the existence of a primary constraint, these actions propagate
   at most three degrees of freedom. See Appendix \ref{app-primary} -- based on 
\cite{Crisostomi:2016czh}  -- for full details.

\smallskip

At the light of these facts, we can use the results we obtained in the previous sections to determine a geometrical
perspective for these particular cases of EST theories.
We  consider a probe brane in AdS bulk, as described in Section \ref{sec-adsdbi}. We
have seen in the previous subsection that the $\kappa_0\to0$ limit of the combination
$\kappa_0\,g^{\mu\alpha} K_{\alpha \nu}$ reads
\be
\lim_{\kappa_0\to0}\,\kappa_0\,g^{\mu\alpha} K_{\alpha \nu}\,=\,\frac{e^{-\pi/\ell}}{ \sqrt{X}}P^\mu_\alpha \,
 \nabla^\alpha \partial_\nu \pi\,=\,\frac{e^{-\pi/\ell}}{ \sqrt{X}}\, Q^\mu_\nu \label{relqmunu1}
\ee
So it is proportional to $Q^\mu_\nu$. (The limit $\ell\to\infty$ corresponds to a brane embedded in a Poincar\'e bulk.) 
On the other hand,
\be
\frac{1}{\kappa_0^3}\,\sqrt{-g}\,=\,\sqrt{-q}\,e^{-3 \pi/\ell}\,\sqrt{\kappa_0^2\,e^{-2 \pi/\ell}+X}
\ee
 This implies that any probe brane action built as a
scalar related to $\kappa_0\,g^{\mu\alpha} K_{\alpha \nu}$ automatically propagates at
most three degrees of freedom in the limit $\kappa_0\to0$, being a special case
of an EST action \eqref{est-a1}. For example, using the results of the previous sections, any action of the form 
($a_i$ are constant parameters)
\be \label{est-tc}
S\,=\,\lim_{\kappa_0\to0}\,\,\frac{{\cal N}}{\kappa_0^3}\,\int d^4 x\,\sqrt{-g}\,\left[1+a_1\,{\rm tr}\left(\kappa_0\,g^{\mu\alpha} K_{\alpha \nu}\right)
+a_2\,{\rm tr}\left(\kappa_0\,g^{\mu\alpha} K_{\alpha \nu}\right)^2+
a_3\,{\rm tr}^2\left(\kappa_0\,g^{\mu\alpha} K_{\alpha \nu}\right)+\dots
 \right]
\ee
belongs to EST theories, and consequently 
propagates
no more than three degrees of freedom. 

\smallskip

Do these actions satisfy some symmetries in certain limits, which can protect their structure against loop corrections,
and lead for example to non renormalization theorems? It depends, and a geometrical approach in terms of a brane
probing an extra dimensional space  can be useful to investigate this question.
 First, let us discuss  the case of gravity {\it not} dynamical. 
 If these theories correspond to super
relativistic ($\kappa_0\to0$) limits of probe brane actions, the existence of symmetries depend on the presence
of isometries in the bulk space time,  as discussed in \cite{Goon:2011uw,Goon:2011qf} and reviewed in the previous sections. However, 
 in general,  actions as eq. \eqref{est-tc} do not admit
a `decoupling limit' around Minkowski space  where gravitational  dynamics can be set to zero and only scalar
dynamics  can  be considered 
 (unless they reduce to beyond Horndeski), since Minkowski space is not
 a solution of the equations. On the other hand,  scalar 
theories can be well defined around some non trivial backgrounds with isometries, and exhibit new symmetries
which are 
 different
from the ones considered here. Second, for the case with dynamical gravity, the same considerations
 of the previous sections apply.  Gravity tends to break all scalar symmetries, but there might be
 cases where such symmetries are broken only in a soft way,
or 
  generalisations
 which promote  the scalar symmetries to full scalar-tensor symmetries.
  We intend to develop these interesting issues in a future analysis.

\section{Discussion}\label{sec-disc}

In this work, we investigated a geometrical approach to degenerate scalar-tensor theories, with the main aim
 to investigate   symmetries and dualities that they satisfy. Using such view point, 
we found a connection between beyond Horndeski (and more  general degenerate scalar-tensor theories of gravity) and  a certain limit of DBI Galileons. 

We started presenting  a perspective on DBI Galileons  based on a determinantal approach. In absence of dynamical gravity,
a particular limit of  DBI Galileons -- which we  called extreme relativistic --  leads to classes of  scalar theories with a
 field dependent symmetry,  that are connected
  by dualities.
   These theories reveal problematic properties when one computes the kinetic terms of  fluctuations around a given background.
   Such problems  can be tamed by weakly breaking the symmetry, by hand 
   or by   coupling the scalar theory to gravity. In the latter case, we 
       showed that a minimal covariantization of  DBI Galileons in the   extreme relativistic
  limit leads  to beyond Horndeski systems, or more in general to degenerate scalar-tensor theories which are consistent despite having equations
  motion of order higher than two.  Our results indicate that  
degenerate scalar-tensor theories can admit a geometrical interpretation in terms of particular limits of  DBI Galileon set-ups, and that
(in absence of dynamical  gravity)  
 they enjoy symmetries which are different from Galileons. Moreover, different special cases
  of beyond Horndeski theories are connected by a duality, in some cases also with
     dynamical gravity. 
 Our results  can be helpful
for assessing the stability properties or understanding the non-perturbative structure of systems based on degenerate scalar-tensor theories.

\smallskip
 Our results can be extended in several  directions. 
 Our geometrical construction of degenerate scalar-tensor theories in terms of branes probing extra dimensions indicates  that special  theories   can be obtained
 when the brane probes specific points in the extra dimensions, where the coefficient in front of the time coordinate vanishes. It would be interesting to 
 examine 
   this
 observation further,  investigating in more general terms scalar-tensor theories obtained   from branes placed
  in  special locations of the embedding space time.  Other possible developments concern symmetries and dualities. 
  We have shown that the scalar symmetry  
  is normally broken when gravity is made dynamical: it would be interesting to find concrete systems
or situations -- specific subclasses of our theories, for example expanded around specific configurations -- where the symmetry breaking can be soft
and controllable, and some of the features
of the symmetry can be maintained. Also, we have studied some special case of duality when gravity is dynamical: it would be interesting to extend the discussion
to study dualities connecting other examples 
 of degenerate scalar-tensor theories.
  Finally, it would be interesting to  study whether  
additional fields can be included in these systems,  still preserving the properties that we determined. We plan to investigate these questions in future studies.

\subsection*{Acknowledgements}

We thank Marco Crisostomi and Kazuya Koyama for many discussions and constructive criticisms, and for collaborating at some stages of the project. We also 
thank Claudia de Rham and Matteo Fasiello for discussions on related topics, and 
Ivonne Zavala for careful reading of the manuscript. 

\begin{appendix}


\section{de Sitter case}\label{app-des}

 {In this Appendix we discuss an alternative, model dependent way to
incorporate a case of beyond Horndeski in a brane world scenario described by an action of the same
determinantal form that we introduced in the main body of the text, e.g. eq. \eqref{ac-dbi-sec}. The
physical effect of the limit that we discuss here is the same as in the case $\kappa\to 0$: to get a
situation where the derivative terms of $\pi$ dominate the induced metric. However, it is important
to emphasize that this alternative limit is formally different from $\kappa \to 0$ and only works for
specific brane/bulk configurations.

By construction, action \eqref{ac-dbi-sec} inherits a global symmetry
from the Killing symmetries of the bulk \cite{Goon:2011uw,Goon:2011qf}, and propagates
the right number of degrees of freedom in the limit to beyond Horndeski.
The explicit form of the scalar symmetry depends on the isometries of the bulk metric
and the geometry of the brane, as explained in \cite{Goon:2011qf}, where maximally
symmetric embeddings are worked out in detail. Among all the possible configurations,
there are two that admit a limit $f(\pi)\to 0$ controlled only by the (A)dS radius,
irrespectively of the value of $\kappa_0$. These two cases are the following:

    \begin{enumerate}
    \item A Minkowski brane embedded in an AdS$_5$ bulk. This is the same geometrical
      configuration that we discussed in section \ref{sec-adsdbi}. Here $f(\pi) =
      e^{-\pi/\ell}$, and the limit $f\to0$ is achieved when $\ell\to0$ and $\pi>0$. Under
      these considerations, the results presented in sec. \ref{sec-adsdbi} are recovered 
      by redefining $\pi = \pi/\kappa_0$ and $\ell = \kappa_0\ell$.
    \item \label{dsds} A  dS$_4$ brane embedded in a  dS$_5$ bulk. Here $f = \ell
      \sin\left(\pi/\ell\right)$ and the limit $f\to0$ corresponds to $\ell\to0$. In
      contrast to the previous case, this limit cannot be related to a limit taken with
      $\kappa_0$ by a redefinition of $\ell$ and $\pi$. We describe this model in detail below.
    \end{enumerate}

The set-up of point \ref{dsds} has the induced metric 
    \begin{equation}
      \label{eq:dsg}
      g_{\mu\nu} = f(\pi)^2 q_{\mu\nu} + \nabla_\mu\pi \nabla_\nu \pi\,, \\
    \end{equation}
    with $f = \ell \sin\left(\frac{\pi}{\ell}\right)$ and $q_{\mu\nu}$
    the metric on the dS$_4$ slices that foliate the dS$_5$ bulk. Indices are raised and lowered
    with $q_{\mu\nu}$. The
    extrinsic curvature $K_{\mu\nu}$ is constructed according to
    \eqref{fextcur}, but replacing $\eta_{\mu\nu}$ with $q_{\mu\nu}$.
    The contraction of the matrix inverse of \eqref{eq:dsg} with the
    extrinsic curvature takes the form
    \begin{equation}
      \label{eq:ginvke}
      g^{\mu\alpha}K_{\alpha\nu} = \frac{\gamma}{f^2}\left(\delta^\mu_\alpha - \frac{\gamma^2}{f^2}\nabla^\mu\pi\nabla_\alpha\pi\right)
\left(-\nabla^\alpha \nabla_\nu \pi + f f' \delta^\alpha_\nu + 2 \frac{f}{f'} \nabla^\alpha\pi \nabla_\nu\pi \right),
    \end{equation}
with $\gamma = f/\sqrt{f^2 + X}$. For this embedding, the action can be put in beyond Horndeski form by taking the
limit $\ell\to0$. To see this, first note that 
\begin{equation}
\lim_{\ell\to0} \left( \delta^\mu_\alpha - \frac{\gamma^2}{f^2}\nabla^\mu\pi\nabla_\alpha\pi \right)\nabla^\alpha \pi = 0\, .
\end{equation}
Now, since $f f' \sim {\cal O}(\ell)$, in the limit $\ell\to0$ the action is dominated by
\begin{equation}
S_{dSG} = \int d^4x \sqrt{-q}\ell^3 \sin^3(\pi/\ell) \sqrt{X} \det\left[\delta^\mu_\nu + \frac{c_1}{\ell \sin{(\pi/\ell)}\sqrt{X} }\left( \delta^\mu_\alpha - \frac{\nabla^\mu\pi\nabla_\alpha\pi}{X} \right)\nabla^\alpha\nabla_\nu \pi  \right],
\end{equation}
which is well-defined if $|c_1/\ell|$ is kept finite when taking the
limit $\ell\to0$. This action belongs to beyond Horndeski, c.f. \eqref{byhact}. The
symmetries of this action -- the limit $\ell\to0$ of the
transformations generated by the Killing vectors of a $dS_4$ brane
embedded in a $dS_5$ bulk derived in \cite{Goon:2011qf} -- are
\begin{eqnarray}
  \label{eq:dsgsym}
  \delta_+ \pi & = & -\cot\left(\frac{\pi}{\ell} \right)  \partial_u \pi, \\
  \delta_- \pi & = & -(u^2 + y^2) \cot\left(\frac{\pi}{\ell} \right)  \partial_u \pi 
                     - 2 u \cot\left(\frac{\pi}{\ell} \right) y^i \partial_i\pi, \\
  \delta_i \pi & = & - y_i \cot\left(\frac{\pi}{\ell} \right)\partial_u \pi - u \cot\left(\frac{\pi}{\ell} \right)\partial_i \pi,
\end{eqnarray}
where $u$ and $y^i$ ($ y^2 = \delta_{ij}y^i y^j,\ i=1,2,3$) are coordinates defined in terms of the bulk coordinates, such that
the induced metric can be written as 
\begin{equation}
ds^2 = d\pi^2 + \ell^2 \sin^2\left(\frac{\pi}{\ell}\right)\left[\frac{1}{u^2}(-du^2 + dy^2)\right].
\end{equation}
These symmetries differ from the symmetries for finite $\ell$ only in
that the Killing vectors have lost their $\partial_\pi$ components,
these components are associated to the part of the symmetry
transformation that does not depend on $\pi$. 
}

\section{Existence of a primary constraint }
\label{app-primary}

In this appendix, we review the main arguments that prove that actions of the form \eqref{est-a1} (which generalise and include Beyond Horndeski systems as discussed
in eq. \eqref{actk01bA}) are free of Ostrogradsky instabilities, and propagate at most three degrees of freedom.  We consider
  the quantities
 $X\,=\,\nabla_\mu \pi \nabla^\mu\pi$ and
 \be
 Q_\mu^\nu\,=\,\left(\delta_\mu^\rho-\frac{\nabla_\mu \pi \nabla^\rho \pi}{X}\right)\,\nabla_{\rho} \nabla^\nu \pi
 \,,
 \ee
and examine arbitrary scalar-tensor  actions (calling $q_{\mu\nu}$ the metric tensor) of the form
\be\label{app-ac1}
S\,=\,\int\,d^4 x\,\sqrt{-q}\,\left[ B_1+ B_2\,Q_\mu^\mu+B_3 \left(Q_\mu^\mu\right)^2+B_4 \left(Q_\mu^\nu \,Q_\nu^\mu\right)+\dots \right]
\,,
\ee
where the $B_i$ are arbitrary functions of $\pi$, $X$. An immediate issue arises:  action \eqref{app-ac1} contains second derivatives of the scalar field. 
 Hence, besides the metric and the scalar $\pi$, actions as \eqref{app-ac1} would seem to  propagate  an additional, fourth mode mode -- related with the scalar velocity $\dot \pi$ -- which is associated with an Ostrogradsky instability.
  In this appendix, we show that this issue does not actually apply for actions  \eqref{app-ac1}:
 there exists a primary constraint which relates
 the dynamics of the scalar velocity with the dynamics of the metric, so to have a system which propagates at most three -- and not four -- degrees of freedom. 

\smallskip

We do so using the geometrical approach introduced by Langlois and Noui, and further developed in \cite{Langlois:2015cwa}; in particular, we review the arguments as presented in \cite{Langlois:2015cwa,Crisostomi:2016czh}.
 We decompose the four dimensional space time in  $3+1$ dimensional slices: we assume there exists a foliation of space time  on $t=const$ hypersurfaces. 
 We can then define on each hypersurface a `time vector' $t^\mu$ as 
\be
t^\mu\,=\,{ N}\,n^\mu+N^\mu
\,,
\ee
with $n^\mu$ the normal, and $N$ and $N^\mu$ respectively the lapse and shift vector. Such time vector determines the time evolution
of the fields involved.  
The $3+1$ decomposition allows us to consider two quantities which further
 characterise the hypersurface geometry:
 \be
 h_{\mu}^\nu\,=\,\delta_\mu^\nu
+n^\nu n_\mu
\,,
 \ee
is the induce metric on the hypersurface; while
\be
K_{\mu\nu}\,=\,\frac{1}{2 N}\left(\dot{h}_{\mu\nu}-\nabla_{(\mu} N_{\nu)}\right)
\,,
\ee
is the hypersurface extrinsic curvature. Here dot indicates the Lie derivative
\be
\dot{h}_{\mu\nu}\,=\,t^\rho\,\nabla_{\rho}\,{h}_{\mu\nu}
\,.
\ee
Instead of using $\nabla_\mu \pi$ in  the scalar-tensor  action, 
it is convenient to  express it in terms of a vector $A_\mu$ defined as
\be\label{app-rel1} 
A_\mu\,=\,\nabla_\mu \pi
\,,
\ee
so that $X\,=\,A_\mu A^\mu$ and
\be
Q_\mu^\nu\,=\,\left(\delta_\mu^\rho-\frac{A_\mu A^\rho }{X}\right)\,\nabla_{\rho}A^\nu
\,. 
\ee
After expressing the action in terms of $A_\mu$, it is easy to `go back' to an expression in terms of $\pi$ only, if one wishes to do so,  by imposing relation \eqref{app-rel1} by means
of a Lagrange multiplier.  The $3+1$ decomposition of space time can be implemented on the vector $A_\mu$ and its covariant derivative
as
\bea
A_\mu&=&-A_\star \,n_\mu+\hat A_\mu
\,,
\\
\nabla_\mu A_\nu&=&D_\mu\,\hat A_\nu
-A_\star\,K_{\mu \nu}+
n_{(\mu}\,\left(K_{\nu) \rho}\,\hat A^\rho-D_{\nu)}A_\star
\right)
n_\mu n_\nu \left( V_\star-\hat A_\rho a^\rho\right)
\,,
\eea
where $(\dots)$ on the index denotes symmetrization (with no numerical factors in front) and 
$a^\rho\,=\,n^\sigma\,\nabla_\sigma n^\rho$ is the acceleration vector. 
We have to consider three quantities which characterise the time flow of the fields described by the action. The first is  time derivative of the metric, conveniently described
by the extrinsic curvature $K_{\mu\nu}$: there are generically two degrees of freedom associated with this quantity (as expected for a spin 2 massless tensor). The second is the time derivative of the scalar, described by $A_\star$ (one dof). The third one is the time derivative of the scalar
time derivative  (one dof), controlled by the quantity
\be
V_\star\,=\,n^\mu A_\mu\,=\,\frac{1}{N}\left(\dot{A}_\star-N^\mu\,\nabla_\mu A_\star\right)
\,.
\ee
Hence action \eqref{app-ac1} propagates 4 dofs, unless there are constraint conditions.  
In what follow, in order to identify the kinetic terms in the action and express everything in terms of   
covariant quantities, it is easier to work  directly with the 
extrinsic curvature $K_{\mu\nu}$ and with $V_\star$, rather than the velocities $\dot h_{\mu\nu}$ and $\dot A_\star$. 
We now show that   constraint conditions exist in the form of primary constraints,   by proving that a linear combination
of the conjugate
momenta
\bea
\pi_\star&=&\frac{1}{\sqrt{-q}}\,\frac{\delta S}{\delta V_\star}
\,,
\\
\pi_\mu^\alpha&=& \frac{1}{\sqrt{-q}}\,\frac{\delta S}{\delta  K_\alpha^\mu K_\star}
  \,,
  \eea
  vanishes. This fact forbids the propagation of a fourth mode.
 
  Using the definition of projection tensor, one has the important relation
  \be \label{imp-rel1}
  A_\star\,P_{\mu}^\nu\,n_\nu\,=\,P_{\mu}^\nu\,\hat A_\nu
\,.
  \ee
  Moreover, 
  \bea
  \frac{\delta Q_\mu^\nu}{\delta V_\star}&=&P_{\mu}^\rho\,n_\rho n^\nu
 \,,
  \\
   \frac{\delta Q_\mu^\nu}{\delta K_\alpha^\beta}\,\hat A_\beta\, \hat A^\alpha
  &=&P_{\mu}^\rho\,\left( -A_\star \hat A_\rho \hat A^\nu+\hat A^2\,n_{\rho}\,\hat A^{\nu}+\hat A^2\,\hat A_{\rho}\,n^{\nu} \right)
 \,.
  \eea
  Using the fact that the action $S$ in eq. \eqref{app-ac1} is a sum of powers of traces of $Q_\mu^\nu$ and its
  powers, as well as relation \eqref{imp-rel1}, one finds the following linear relation among conjugate momenta
 (with the notation $\approx$ we mean weak inequality, that is inequality in the phase space of constraints)
  \be
  A_\star \left( 2 \hat A^2-A_\star^2\right)
  \,\pi_\star-\hat A_\rho \,\hat A^\sigma\,\pi_\sigma^\rho\,\approx\,0
 \,.
  \ee
  Hence, there exists a primary constraint
  which forbids the propagation of a fourth mode for  theories
  described by an action \eqref{app-ac1}.
   The theories that we are investigate propagate at most three degrees of freedom. 

\end{appendix}

\end{document}

%% file: branepipos.pdf_t
\begin{picture}(0,0)%
\includegraphics{branepipos.pdf}%
\end{picture}%
\setlength{\unitlength}{4144sp}%
\begingroup\makeatletter\ifx\SetFigFont\undefined%
\gdef\SetFigFont#1#2#3#4#5{%
  \reset@font\fontsize{#1}{#2pt}%
  \fontfamily{#3}\fontseries{#4}\fontshape{#5}%
  \selectfont}%
\fi\endgroup%
\begin{picture}(7647,3182)(1916,-7733)
\put(3622,-7439){\makebox(0,0)[lb]{\smash{{\SetFigFont{17}{20.4}{\familydefault}{\mddefault}{\updefault}{\color[rgb]{0,0,0}$M_4$}%
}}}}
\put(6771,-7059){\makebox(0,0)[lb]{\smash{{\SetFigFont{17}{20.4}{\familydefault}{\mddefault}{\updefault}{\color[rgb]{0,0,0}$\pi=0$}%
}}}}
\put(6880,-4887){\makebox(0,0)[lb]{\smash{{\SetFigFont{17}{20.4}{\familydefault}{\mddefault}{\updefault}{\color[rgb]{0,0,0}$\pi(x^\mu)$}%
}}}}
\end{picture}%